\documentclass[12pt]{article}

\usepackage[fleqn]{amsmath}
\usepackage{amsfonts}
\usepackage{amssymb}
\usepackage{epsfig}
\usepackage{geometry}
\usepackage{ctable}
\usepackage[all,cmtip]{xy}

\makeatletter

\renewcommand\thesection {\@arabic\c@section}
\renewcommand\thesubsection   {\thesection.\@arabic\c@subsection}
\renewcommand\thesubsubsection{\thesubsection .\@arabic\c@subsubsection}
\renewcommand\theparagraph    {\thesubsubsection.\@arabic\c@paragraph}
\renewcommand\section{\@startsection {section}{1}{\z@}%
                                   {-3.5ex \@plus -1ex \@minus -.2ex}%
                                   {1.9ex \@plus.2ex}%
                                   {\normalfont\large\bfseries\centering}}
\renewcommand\subsection{\@startsection{subsection}{2}{\z@}%
                                     {-2ex\@plus -1ex \@minus -.2ex}%
                                     {1.2ex \@plus .2ex}%
                                  {\normalfont\normalsize\bfseries\centering}}
\renewcommand\subsubsection{\@startsection{subsubsection}{3}{\z@}%
                                     {-2ex\@plus -1ex \@minus -.2ex}%
                                     {.5ex \@plus .2ex}%
                                     {\normalfont\normalsize\em}}
\renewcommand\paragraph{\@startsection{paragraph}{4}{\z@}%
                                    {3.25ex \@plus1ex \@minus.2ex}%
                                    {-1em}%
                                    {\normalfont\normalsize\em}}
\makeatother

\newcounter{subequation}
        \newenvironment{subequation}%
        {\addtocounter{equation}{-1}%
        \stepcounter{subequation}%
        \begin{equation}}%
        {\end{equation}%
}

\newcommand{\beq}{\begin{equation}}
\newcommand{\eeq}{\end{equation}}
\newcommand{\bseq}{\begin{subequation}}
\newcommand{\eseq}{\end{subequation}}
\newcommand{\bea}{\begin{eqnarray}}
\newcommand{\eea}{\end{eqnarray}}

\newcommand\hfrac[2]{{#1\big/#2}}

\newcommand{\cO}{{\cal O}}

\newcommand{\cR}{{\cal R}} % Radiation Reaction term

\newcommand{\Bset}{{\mathbb B}}

\newcommand{\Rset}{{\mathbb R}}

\newcommand{\Csp}{{\mathfrak{C}}}

\newcommand{\Lsp}{{\mathfrak{L}}}

\newcommand{\Psp}{{\mathfrak{P}}}

\newcommand{\QED}{$\quad$\textrm{Q.E.D.}}

\newcommand{\txtcinv}{{\textstyle{\frac{1}{c}}}}

\newcommand{\supN}{\scriptscriptstyle{(N)}}
\newcommand{\supNb}{\scriptscriptstyle{(N_\beta)}}

\newcommand{\dd}{{\mathrm{d}}}

\newcommand{\pdt}{{\partial_t^{\phantom{0}}}}

\numberwithin{equation}{section}

\newtheorem{thm}{Theorem}[section]

\newtheorem{rem}[thm]{\sc Remark}
\newtheorem{obs}[thm]{\sc Observation}

\newcommand{\Cdot}{{\,\cdot\,}}
\newcommand{\Cddot}{{\,\cdot\,,\,\cdot\,}}

\newcommand{\veps}{\varepsilon}

\newcommand{\drm}{\mathrm{d}}

\newcommand{\Ddt}{\frac{\drm\phantom{s}}{\drm t}}

\newcommand{\pdp}{\partial_{\pV}}
\newcommand{\pds}{\partial_{\sV}}
\newcommand{\pddt}{\frac{\partial\phantom{t}}{\partial t}}
\newcommand{\tpddt}{{\textstyle{\frac{\partial\phantom{t}}{\partial t}}}}

\newcommand{\tpddct}{{\textstyle{\frac{1}{c}\frac{\partial\phantom{t}}{\partial t}}}}

\newcommand{\ONE}{{\boldsymbol{1}}}

\newcommand{\refeq}[1]{(\ref{#1})}

\newcommand{\vect}[1] {\boldsymbol{{ #1}} }

\newcommand{\aV}{\vect{a}}              % 3-acceleration

\newcommand{\fV}{\vect{f}}              % 3-force density vector field
\newcommand{\jV}{{\vect{j}}}		% 3-current density
\newcommand{\nV}{{\vect{n}}}		% 3-unit vector
\newcommand{\pV}{{\vect{p}}}            % momentum 3-vector 
\newcommand{\qV}{{\vect{q}}}            % 3-position of particle
\newcommand{\olqV}{{\overline{\vect{q}}}}            % 3-position of particle
\newcommand{\sV}{{\vect{s}}}            % position 3-vector
\newcommand{\vV}{{\vect{v}}}            % 3-velocity of particle

\newcommand{\NullV}{\vect{0}}

\newcommand{\BV}{\pmb{{\cal B}}}

\newcommand{\DV}{\pmb{{\cal D}}}
\newcommand{\EV}{\pmb{{\cal E}}}

\newcommand{\HV}{\pmb{{\cal H}}}

\newcommand{\PiV}{\boldsymbol{\Pi}}
\newcommand{\nab}{\partial_{\sV}}

\newcommand{\abs}[1]{\big\vert #1 \big\vert}
\newcommand{\Abs}[1]{\left| #1 \right|}

\newcommand\Lip{\mathrm{Lip}}

\newcommand\dist[1]{{{\mathrm{dist}}\left( #1\right)}}        % input is a,b

\newcommand{\ul}[1]{\underline #1 }
\newcommand{\oli}[1]{\overline #1 }
\newcommand{\uli}[1]{\underline #1 }

\renewcommand{\leq}{\leqslant}
\renewcommand{\geq}{\geqslant}

\newcommand{\crprd}{{\boldsymbol\times}}

\newcommand{\kB}{k_{\mbox{\tiny{B}}}}
\begin{document}

\title{Microscopic foundations of kinetic plasma theory:\\
       The relativistic Vlasov--Maxwell equations and \\
       their radiation-reaction-corrected generalization\vspace{-.5truecm}}
% \titlerunning{Vlasov--Maxwell limit}

\author{Y. Elskens$^1$ and M.K-H. Kiessling$^2$ \\
  $^1$ \footnotesize{Aix-Marseille universit\'e, UMR 7345 CNRS,}\vspace{-.1truecm}\\
       \footnotesize{case 322, campus Saint-J\'er\^ome,
       F-13397 Marseille cedex 13}\vspace{-.1truecm}\\
       \footnotesize{email : yves.elskens@univ-amu.fr}\\
  $^2$ \footnotesize{Department of Mathematics, Rutgers, The State University of New Jersey,}\vspace{-.1truecm}\\
       \footnotesize{110 Frelinghuysen Rd., Piscataway, NJ 08854}\vspace{-.1truecm}\\
       \footnotesize{email : miki@math.rutgers.edu} \vspace{-.5truecm}}
\date{\footnotesize{Printed \today}}
\maketitle
\vspace{-1.3truecm}
\begin{abstract}\noindent
 It is argued that the relativistic Vlasov--Maxwell equations of the kinetic theory of plasma
% and a radiation-reaction corrected generalization thereof, 
approximately describe a relativistic 
system of $N$ charged point particles interacting with the electromagnetic Maxwell fields in a 
Bopp--Land\'e--Thomas--Podolsky (BLTP) vacuum, provided the microscopic dynamics lasts long enough. 
 The purpose of this work is not to supply an entirely rigorous vindication, but to lay down 
a conceptual road map for the microscopic foundations of the kinetic theory of special-relativistic 
plasma, and to emphasize that a rigorous derivation seems feasible.
 Rather than working with a BBGKY-type hierarchy of $n$-point marginal probability measures, the 
approach proposed in this paper works with the distributional PDE of the actual empirical 1-point measure, which 
involves the actual empirical 2-point measure in a convolution term.
 The approximation of the empirical 1-point measure by a continuum density, and of the empirical
2-point measure by a (tensor) product of this continuum density with itself, yields a finite-$N$ 
Vlasov-like set of kinetic equations which includes radiation-reaction and nontrivial finite-$N$ 
corrections to the  Vlasov--Maxwell--BLTP model. 
 The finite-$N$ corrections formally vanish in a mathematical scaling limit $N\to\infty$ in which charges $\propto 1/\surd{N}$.
 The radiation-reaction term vanishes in this limit, too.
 The subsequent formal limit sending Bopp's parameter $\varkappa\to\infty$ yields the Vlasov--Maxwell model.
\end{abstract}

\vfill \hrule
\smallskip
 \noindent {\footnotesize{\copyright{2020} The authors. Reproduction of this article, in its entirety,
for noncommercial purposes is permitted.}}\vspace{-.5truecm}

\newpage

%===========================================================
\section{Introduction}\vspace{-.3truecm}
\label{intro}
%===========================================================

\noindent
 An estimated $99.99\%$ of the visible matter in the universe is fully ionized plasma, matter so hot
that the negatively charged electrons and several species of positively charged nuclei are not in any
bound state.
 This hot state of matter is essentially described by a classical, so-called ``collisionless kinetic theory'' ---
an unlucky name for a plasma theory in which a Balescu--Guernsey--Lenard-type dissipation operator is neglected.
 The stellar-dynamical ``collisionless kinetic theory'' for a system of $N$ stars \cite{Jeans}, treated 
as point particles interacting with Newton's gravity, serving as a template, Vlasov \cite{LuchinaVlasov, VlasovBOOK}
proposed a kinetic theory of plasma in which each particle species $\sigma$ is assigned, at time $t\geq 0$, a  
continuum density $N_\sigma f_\sigma(t,\sV,\pV) \geq 0$ at the kinematical momentum $\pV$ in the co-tangent space 
$\Rset^3_\sV$ of the location $\sV$ in  physical space $\Rset^3$.
  Each  $f_\sigma \in \Csp^1(\Rset,(\Psp_1\cap\Csp^1)(\Rset^6))$,
where $\Psp_1\cap\Csp^1$ denotes the normalized measures which have densities in $\Csp^1$ and finite first moments,
satisfies the incompressible-transport equation
\begin{equation}
\pdt f_\sigma(t,\sV,\pV)
+ \vV_\sigma \cdot\pds f_\sigma(t,\sV,\pV)
+ e_\sigma
\left(\EV(t,\sV) +
\txtcinv
\vV_\sigma\times \BV(t,\sV)\right)\cdot\pdp f_\sigma(t,\sV,\pV)
= 0\,,
\label{eq:rVMfEQs}
\end{equation}
where $c$ is the vacuum speed of light, $e_\sigma$ and $m_\sigma$ are 
electric charge and rest mass of a particle of species $\sigma$, whose
velocity $\vV_\sigma=\vV_\sigma(\pV)$ is related to the generic $\pV$ variable by
\begin{equation}
  \vV_\sigma(\pV) = \hfrac{c\pV}{\sqrt{m_\sigma^2 c^2 + \abs{\pV}{}^2}}.
  \label{EINSTEINvOFp}
\end{equation}
  In \refeq{eq:rVMfEQs}, $\EV(t,\sV)$ and $\BV(t,\sV)$ are self-consistent
electric and magnetic fields at the space point $\sV$ at time $t$.
 These are not the true microscopic fields but are defined at the same (reduced) level of resolution as the distributions
of particle positions and momenta, i.e.\ classical fields
$\EV(\Cddot)\in \Csp^1(\Rset,{(\Lsp^2\cap\Csp^1)}(\Rset^3))$ and $\BV(\Cddot)\in \Csp^1(\Rset,(\Lsp^2\cap\Csp^1)(\Rset^3))$
satisfying the system of field equations (in Heaviside--Lorentz units, following \cite{SpohnBOOKb})
\begin{eqnarray}
        \pdt{{\BV}(t,\sV)} + c \pds\times{\EV}(t,\sV)
  \!\!&=&\!\!
        \NullV  \, , 
 \phantom{{\textstyle\sum\limits_\sigma^{}} N_\sigma e_\sigma \int_{{\Rset}^3}f_t^\sigma (\sV,\pV) bla} 
  \label{eq:rVMrotE} \\
 \pds\cdot {\BV}(t,\sV)
  \!\!&=&\!\!
        0\, , \phantom{{\textstyle\sum\limits_\sigma^{}} N_\sigma e_\sigma \int_{{\Rset}^3}f_t^\sigma (\sV,\pV)bla}
  \label{eq:rVMdivB}\\
\qquad  -\pdt{{\DV}(t,\sV)} + c\pds\times{\HV}(t,\sV) 
  \!\!&=&\!\!
           {\textstyle\sum\limits_\sigma^{}} N_\sigma e_\sigma
     \int_{{\Rset}^3} \vV_\sigma(\pV) f_\sigma (t,\sV,\pV)\dd^3{p}\,, 
  \label{eq:rVMdotD}  \\
 \qquad        \pds\cdot{\DV}(t,\sV)
  \!\!&=&\!\!
        %  4 \pi
        {\textstyle\sum\limits_\sigma^{}} N_\sigma e_\sigma \int_{{\Rset}^3}f_\sigma (t,\sV,\pV)\dd^3{p},
  \label{eq:rVMdivD}
\end{eqnarray}
together with Maxwell's ``law of the pure ether''
\begin{eqnarray}
\qquad {\DV}(t,\sV) 
  =
{\EV}(t,\sV) \qquad \&\ \qquad 
 {\HV}(t,\sV)
  =
 {\BV}(t,\sV).
  \label{eq:rVMetherM}
\end{eqnarray}

  This set of equations is Lorentz covariant, though not written manifestly so:
the dynamical variables are defined in a Lorentz frame with time \&\ space coordinates $(t,\sV)$.

 Vlasov's kinetic equations find applications in the relativistic theory of so-called collective phenomena in
astrophysical plasma, see \cite{SchindlerJanicke, Janicke, Otto, Neukirch, SchlickeiserKneller}; also see
the books \cite{Schlickeiser,Schindler,Somov,VereshchaginAksenov}.
 They also find applications in the relativistic theory of laser-plasma interactions, see \cite{InglebertETal}. 
 For a recent review, see \cite{PalmrothETal}, and for a general relativistic setting, see \cite{EhlersA, EhlersB, Bernstein}.

  Known in the mathematical literature as (relativistic) Vlasov--Maxwell equations,
their Cauchy problem has been settled for small initial data, and for large data under the additional
assumption that no singularities occur near the light cone,
see~\cite{GlasseyStraussA,GlasseyStraussB,GlasseySchaefferB,KlaSta,GlasseyPankavichSchaeffer};
related rigorous treatments of relativistic Vlasov equations are
\cite{HorstHABIL, GlasseyStraussC, diPLioB, ReinA, BoGoPaI, BoGoPaII, MikiShadi}. 
 Also the work of Mouhot and Villani \cite{MouhotVillaniA} on Landau damping in non-relativistic plasma 
has been generalized to the relativistic Vlasov--Maxwell system, see \cite{BOJYa, BOJYb}.
Most recently, an Onsager type conjecture on conservation of energy and entropies of weak solutions was established 
\cite{Bardos19}. 
 
  An important feature of Vlasov's set of kinetic equations is that they are time-reversal invariant, and in fact preserve 
the Boltzmann entropy functional 
\begin{equation}
S(\{f_\sigma\}) = - \sum_\sigma \kB\int_{\Rset^6} f_\sigma (t,\sV,\pV)\ln\left(h^3 f_\sigma (t,\sV,\pV)\right)\dd^3{p}\dd^3{s}.
\end{equation}
 In a certain very concrete sense, the Vlasov kinetic equations are for a plasma what the kinetic (free streaming) 
equations of an ideal gas are for a dilute gas of
short-range interacting neutral particles: they govern the dynamics only over very short time scales. 
 To go beyond these short time scales, the kinetic equations have to be corrected, which in the non-relativistic regime 
means including pair scattering operators, for historical reasons usually misnamed as ``collision operators:'' 
the Boltzmann pair-collision operator in case of a dilute gas of short-range interacting neutral particles, 
and a Balescu--Guernsey--Lenard type operator (similar to a Boltzmann pair-collision operator except that 
its scattering cross section is a nonlinear functional of the 1-point density function $f$ itself) in case of a fully 
ionized plasma.
 These dissipation operators still involve only the 1-point density functions; yet they also involve a two-particle kernel 
which encodes the pair interaction between particles at the fundamental level.

 A plasma in the relativistic regime radiates a lot of electromagnetic energy and thus the relativistic kinetic equations
describing it should exhibit another dissipation term, due to the radiation-reaction on the 
motion of individual particles which add up to a bulk effect in a large $N$ plasma.
 This in turn is a sum of purely individual dissipation terms and is believed to be more and more important the
hotter (read higher energetic) the plasma is --- note that in this really hot regime the dissipation due to the
Balescu--Guernsey--Lenard-type scattering operator is expected to be negligible, by comparison. 
 Radiation-reaction corrections to the Vlasov--Maxwell system, based on the works of Dirac \cite{DiracA} and
Landau--Lifshitz \cite{LandauLifshitz},  have been proposed in \cite{Hakim68,HakimBOOK}, \cite{Kuzmenkov} and \cite{NBGJ}, 
and an $H$-theorem proved for them in \cite{Hakim68} and \cite{BurtonNobleA}.
 For mathematically rigorous studies of a radiation-reaction corrected Vlasov--Poisson system, see 
\cite{KunzeRendallA, KunzeRendallB} and \cite{Bauer}, and also references therein.

 The microscopic foundations of the Vlasov--Maxwell equations 
\refeq{eq:rVMfEQs}--\refeq{eq:rVMetherM}, and their radiation-reaction corrected generalizations in 
\cite{Kuzmenkov} and \cite{NBGJ},  are still in a lamentable state of affairs.
 These equations are meant to be a continuum approximation to the electrodynamics of
$N=\sum_\sigma N_\sigma$ classical point charges and their joint electromagnetic field, 
when viewed ``coarse-grained'' \cite{Ichimaru,LL10,GoldstonRutherford}.
 In the Vlasov approximation, the point charges move like \emph{test particles}
according to 
\begin{eqnarray}
  \dot{\qV}^{\sigma}(t)
  & = &
\hfrac{c\pV^\sigma(t)}{\sqrt{m_\sigma^2 c^2 +|\pV^\sigma(t)|{}^2}}\,,
  \label{CHqDOT}
  \\
  \dot{\pV}^{\sigma}(t)
  & = &
  e_\sigma
  \Big(\EV\big(t,\qV^\sigma(t)\big) + \txtcinv\dot{\qV}^\sigma(t) \times \BV\big(t,\qV^\sigma(t)\big)\Big) 
\,,
\label{CHpDOT}
\end{eqnarray}
the characteristic system of \refeq{eq:rVMfEQs}, where $\EV$ and $\BV$ are \emph{regular} field solutions 
of~the~Vlasov PDE system with \emph{regular} density functions $f^\sigma$.
   Regrettably, though the issue has been raised as an important challenge long ago \cite{MontgomeryTidman}, 
the conventional plasma physics literature \cite{KrallTrivelpiece,Klimontovich,Nicholson,Piel}
furnishes a putatively underlying atomistic system by simply ``atomizing'' the density functions $f^\sigma_t$.
  Yet, replacing each $f_\sigma(t,\sV,\pV)$  by a normalized empirical  measure
$\frac{1}{N_\sigma} \sum_{k=1}^{N_\sigma} \delta(\pV-\pV_k^\sigma(t))
  \delta(\sV-\qV_k^\sigma(t))$ of $N_\sigma$ point particles with positions $\qV_k^{\sigma}(t)$
and momenta $\pV_k^{\sigma}(t)$ turns the characteristic system into
\begin{eqnarray}
  \dot{\qV}_k^{\sigma}(t)
  & = &
  \hfrac{c \pV_k^{\sigma}(t) }
  { \sqrt{m_\sigma^2 c^2 +\bigl|\pV_k^{\sigma}(t)\bigr|{}^2}}\,,
  \label{qkDOTalpha}
  \\
  \dot{\pV}_k^{\sigma}(t)
  & = &
 e_\sigma\Big(\EV^{\supN}\big(t,\qV^\sigma(t)\big)+\txtcinv\dot{\qV}_k^\sigma(t)\crprd\BV^{\supN}\big(t,\qV^\sigma(t)\big)\Big)\,,
  \label{pkDOTalpha}
  \\
        \txtcinv\pdt{{\BV}^{\supN}(t,\sV)}
  \!\!&=&\!\!
        - \pds\times{\EV}^{\supN}(t,\sV)
  \, ,
  \label{eq:MLrotE}
  \\
         \txtcinv\pdt{{\EV}^{\supN}(t,\sV)}
  \!\!&=&\!\!
  \ \pds\times{\BV}^{\supN}(t,\sV)    - 
  \textstyle\sum\limits_\sigma^{}  e_\sigma
    \textstyle\sum\limits_{k=1}^{N_\sigma} 
\txtcinv\dot{\qV}_k^{\sigma}(t)\delta_{\qV_k^{\sigma}(t)}(\sV)\,,
  \label{eq:MLrotB}
  \\
  \pds\cdot {\BV}^{\supN}(t,\sV)
  \!\!&=&\!\!
        0\, ,
  \label{eq:MLdivB}
  \\
  \pds\cdot{\EV}^{\supN}(t,\sV)
  \!\!&=&\!\!
       % 4 \pi
        \textstyle\sum\limits_\sigma^{} e_\sigma \textstyle\sum\limits_{k=1}^{N_\sigma} \delta_{\qV_k^{\sigma}(t)}(\sV),
\label{eq:MLdivE}
\end{eqnarray}
 the ``symbolic'' Lorentz equations for classical point charges interacting with Maxwell--Lorentz fields.
The reason for why these Lorentz equations are, at best, only of a symbolic
character is well known: Lorentz electrodynamics with point charges is plagued by infinities!

 More precisely, no distributional solution $\EV^{\supN}\!,\,\BV^{\supN}$ to
\refeq{eq:MLrotE}---\refeq{eq:MLdivE} for given particle trajectories is in
$\Lsp^2(\Bset)$ for any open ball $\Bset$ containing the location of a point charge.
 So each point charge is surrounded by an infinite field energy and acquires an infinite
inertia (via Einstein's $E=mc^2$ \cite{einsteinB}); furthermore, the expression for the Lorentz force \cite{LorentzFORCE} at 
the right-hand side of \refeq{pkDOTalpha} is not well-defined: it is ``infinite in all directions'' \cite{Kie12}; furthermore,
typically a singularity propagates along the initial forward light-cone of each and every particle, which
terminates the motion as soon as a particle meets the initial forward light-cone of another one \cite{DeckertHartenstein}.

 Infinite self-interaction terms are encountered also if one applies the atomization
to the strictly non-relativistic Vlasov--Poisson equations.
 But, in this case, the self-interactions are not dynamical, and simply discarding
them formally yields the locally well-posed and consistent Newtonian $N$-body system
for which, in turn, the non-relativistic Vlasov--Poisson evolution \cite{PfaffDISS, Pfaff, SchaefferB}
is expected to be an asymptotically exact approximation when $N\to\infty$; see \cite{BalescuBOOK1,Balescu75} for good 
theoretical physics and \cite{Jab14, Kie14, LazaroviciPickl, Serfaty} for rigorous mathematics.
 For regularized forces, the rigorous derivation of the non-relativistic Vlasov kinetic equations
has been carried out completely; see \cite{NeunzertA,BraunHepp,Dobru,NeunzertB,SpohnBOOKa}.

 However, in the relativistic theory such a formal omission of the self-interaction
terms is not justified because of the dynamical radiation-reaction, claims to the contrary in the plasma
physics literature \cite{KrallTrivelpiece,Klimontovich,Nicholson} notwithstanding.
 Neither the Vlasov--Maxwell system, nor its radiation-reaction corrected generalization,
can be justified using an inconsistent or ill-defined microscopic system of equations.

 Recently, the microscopic foundations of relativistic classical electromagnetic theory with point charges
were formulated in \cite{Kie19} for a class of admissible \emph{electromagnetic vacuum laws}.
 In \cite{KTZonBLTP} a proof of local well-posedness of the joint initial value problem for the fields and point charges 
in the electromagnetic BLTP vacuum was established; see also \cite{VuMaria}.
 The BLTP vacuum is named after 
Bopp \cite{BoppA,BoppB}, Land\'e \&\ Thomas \cite{LandeThomas}, and Podolsky \cite{Podolsky}, 
who proposed the vacuum law
\begin{alignat}{1}
        \DV(t,\sV) 
=
        \left(1  + \varkappa^{-2}\square\,\right) \EV(t,\sV) \qquad \&\ \qquad
        \HV(t,\sV)  
= 
       \left(1  + \varkappa^{-2}\square\,\right) \BV(t,\sV) \, ;
\label{eq:BLTPvaclaw}
\end{alignat}
here, $\square \equiv c^{-2}\partial_t^2 -\Delta$  is the classical wave operator, and
$\varkappa$ is ``Bopp's reciprocal length'' parameter \cite{BoppA}.
 In the singular limit $\varkappa\to\infty$, the BLTP law formally yields Maxwell's law \refeq{eq:rVMetherM}.

 In the present paper, we propose that the pertinent  Vlasov--Maxwell--BLTP equations 
\refeq{eq:rVMfEQs}--\refeq{eq:rVMdivD} \&\ \refeq{eq:BLTPvaclaw} are 
an asymptotically exact approximation to BLTP electrodynamics when $N\to\infty$, provided the
microscopic dynamics lasts long enough. 
 Asymptotic exactness involves a mathematical rescaling of charges and fields ---
 Spoiler Alert: There seems to be \emph{no} physically realizable  scaling with $N$ (i.e.\  keeping all physical constants
unscaled) for which the Vlasov--MBLTP system is asymptotically exact as $N\to\infty$ in the BLTP $N$-body plus field electrodynamics.

 We emphasize that the strategy of naive atomization will \emph{not} produce a viable microscopic 
dynamical theory of $N$ point charges and their Maxwell--BLTP (MBLTP) fields:
 once again the formal Lorentz force of the joint electromagnetic field on a point charge is ill-defined;
while it no longer is ``infinite in all directions,'' it still ``points in all directions,'' and with ``varying
amplitude.''
 It is precisely this problem which was recently overcome in \cite{Kie19}. 

 In the next section, we recall the proper defining equations of BLTP electrodynamics with $N$ point charges.
 In section \ref{sec:VLASOVapprox}, we explain their Vlasov approximation, featuring also a radiation-reaction-corrected
Vlasov--Maxwell--BLTP model.
 We collect some already proven facts and highlight the types of estimates which are still needed to complete the proof. 
 The crucial issue is the one of the time scales. 
 We do not yet know whether BLTP electrodynamics with $N$ point charges is typically globally well-posed or whether finite-time 
singularities will typically form, conceivably through radiative inspiraling of two oppositely charged point particles.
 Even if finite-time blow-up happens typically, the question is how long it will take until a singularity forms. 
 If the time scale for singularity formation is longer than the shortest kinetic theory time scale, the Vlasov time scale, 
then the rigorous derivation of the radiation-reaction corrected Vlasov--Maxwell--BLTP equations from BLTP electrodynamics 
with point charges should be feasible.

 The  Vlasov--Maxwell equations are then conceivably obtained from these radiation-reaction corrected 
Vlasov--Maxwell--BLTP equations on sufficiently short time scales where radiation reaction should be negligible 
(there is such a time scale), and then by taking the subsequent limit in which the BLTP vacuum law \refeq{eq:BLTPvaclaw} 
reduces to Maxwell's vacuum law \refeq{eq:rVMetherM}.
 This is formally obvious but still needs to be proved; see section 4.
\vfill

\centerline{***}

\noindent
 We are honored and pleased to celebrate Joel Lebowitz' 90-th birthday, and his many years of invaluable service to the 
statistical physics community, with this contribution to the microscopic foundations of relativistic kinetic plasma theory, dedicated to Joel.

\newpage

%===========================================================
\section{BLTP electrodynamics with $N$ point charges}\vspace{-5pt}\label{sec:BLTPdyn}
\label{VMBLTP}
%===========================================================

 In this section, we first summarize BLTP electrodynamics in a nutshell.
 BLTP electrodynamics, as defined in \cite{Kie19} and \cite{KTZonBLTP}, couples two sets of equations:
the microscopic MBLTP field equations, and the relativistic version of Newton's equations of motion with 
Poincar\'e's definition of the electromagnetic force. 

%%%%%%%%%%%%%%%%%%%%%%%%%%%%%%%%%%%%%%%%%%%%%%%%%%%%%%%%%%%%%%
%%%%%%%%%%%%%%%%%%%%%%%%%%%%%%%%%%%%%%%%%%%%%%%%%%%%%%%%%%%%%%
           \subsection{\hspace{-10pt} The Maxwell--Bopp--Land\'e--Thomas--Podolsky field theory}\label{sec:MBLTPeqns}
%%%%%%%%%%%%%%%%%%%%%%%%%%%%%%%%%%%%%%%%%%%%%%%%%%%%%%%%%%%%%%
%%%%%%%%%%%%%%%%%%%%%%%%%%%%%%%%%%%%%%%%%%%%%%%%%%%%%%%%%%%%%%
%
 The MBLTP field equations for $N$ point charge sources comprise the inhomogeneous
Maxwell equations
\begin{alignat}{1}
\textstyle
- \pdt{\DV^{\supN}(t,\sV)}         + c\nab\crprd\HV^{\supN}(t,\sV)  
&= 
    \textstyle\sum\limits_\sigma e_{\sigma}
\textstyle{\sum\limits_{\ 1\leq n\leq N_\sigma}}  \delta_{\qV_n^\sigma(t)}(\sV){\vV_{\!n}^\sigma}(t)\, ,
\label{eq:MdotD}
\\
        \nab\cdot\DV^{\supN}(t,\sV)  
&=
    \textstyle\sum\limits_\sigma e_{\sigma} 
        \textstyle{\sum\limits_{\ 1\leq n\leq N_\sigma}} \delta_{\qV_n^\sigma(t)}(\sV)\, ,
\label{eq:MdivD}
\end{alignat}
and the homogeneous Maxwell equations,
\begin{alignat}{1}
\hspace{.5truecm}
\textstyle
\pdt{\BV^{\supN}(t,\sV)}        + c \nab\crprd\EV^{\supN}(t,\sV) 
&= \label{eq:MdotB}
\NullV
\, ,
\\
\hspace{.5truecm}
        \nab\cdot \BV^{\supN}(t,\sV)  
&= \label{eq:MdivB}
        0\, ,
\end{alignat}
and the two vector equations of the BLTP law of the electromagnetic vacuum,
\begin{alignat}{1}
        \HV^{\supN}(t,\sV)  
&= \label{eq:BLTPlawBandH}
       \left(1  + \varkappa^{-2}\square\,\right) \BV^{\supN}(t,\sV) \, ,
\\
        \DV^{\supN}(t,\sV) 
&=
        \left(1  + \varkappa^{-2}\square\,\right) \EV^{\supN}(t,\sV) \, .
\label{eq:BLTPlawEandD}
\end{alignat}

 To solve the MBLTP field equations as an initial value problem, given the particle motions,
requires field initial data $(\BV^{\supN}, \DV^{\supN}, \EV^{\supN}, \dot\EV^{\supN})(0,\,.\,)$,
with $(\BV^{\supN}, \DV^{\supN})(0,\,.\,)$ constrained by \refeq{eq:MdivD} and \refeq{eq:MdivB}.

 The solution to the MBLTP field equations is given as follows.
 Let $\ONE^{}_\Omega$ denote the indicator function of the set $\Omega\subset\Rset^3$.
 Then
\begin{equation}\label{eq:KGkernel}
G^\varkappa_t(\sV) := 
\frac{1}{4\pi}
\Big(
\frac{\delta_{ct}(|\sV|)}{|\sV|} - \varkappa\frac{J_1(\varkappa\sqrt{c^2t^2 - |\sV|^2})}{\sqrt{c^2t^2 - |\sV|^2}} \ONE^{}_{\{|\sV| 
\leq ct \}} \Big) % \ {\bf CHECK\ SIGN} % DONE !
\end{equation}
is the fundamental solution at $t>0$ of the Klein--Gordon equation $(\square +\varkappa^2) u(t,\sV)=0$ with initial data 
$u(0,\sV)\equiv 0$ and $(\frac1c\pdt u(t,\sV))|_{t=0}=\delta(|\sV|)$.
 Note that $G^0_t(\sV)$ is the fundamental solution at $t>0$ of the wave equation $\square u(t,\sV) =0$ with the same initial data.
 Let  $(f*g)(\sV) : = \int_{\Rset^3} f(\sV')g(\sV-\sV')\drm^3{s'}$.
 Then
\begin{alignat}{2}\label{eq:HforN}
\hspace{-.5truecm}
{\HV^{\supN}(t,\sV)} = & 
- \big(G^0_t* \nab\crprd\DV^{\supN}|_{t=0} \big)(\sV) \cr 
&
+ \big(\tpddct G^0_t * 
\big[\left(1  - \tfrac{1}{\varkappa^2}\Delta\,\right) \BV^{\supN}|_{t=0} 
- \tfrac{1}{c\varkappa^2}\nab\crprd(\tpddt\EV^{\supN})|_{t=0}\big]\big)(\sV) \hspace{-3truecm}
\cr 
&
- {\textstyle{\sum\limits_{\sigma} e_\sigma^{}} 
\textstyle{\sum\limits_{1\leq n\leq N_\sigma}} }
\int_0^t \Big(G^0_{t-t'}* \left[{\vV_{\!n}^\sigma}(t')\crprd \nab\delta_{\qV_n^\sigma(t')}\right]\Big)(\sV)\drm{t'} , 
\end{alignat}
\begin{alignat}{1}\label{eq:DforN}
\hspace{-.5truecm}
\DV^{\supN}(t,\sV)
=  
\; \DV^{\supN}(0,\sV) + \int_0^t \Big( c\nab\crprd\HV^{\supN}(t',\sV)  - 
\textstyle{\sum\limits_{\sigma} e_\sigma^{} }
\textstyle{\sum\limits_{1\leq n\leq N_\sigma}} 
\delta_{\qV_n^\sigma(t')}(\sV){\vV_{\!n}^\sigma}(t')\Big)\drm{t'}\, .
\end{alignat}

 Having $\DV^{\supN}$ and $\HV^{\supN}$, the remaining field solutions are given by 
\begin{alignat}{1}
\hspace{-1truecm}
{\BV^{\supN}(t,\sV)} = % 
&  \label{eq:BsolKG} 
\big(\tpddct G^\varkappa_t * \BV^{\supN}|_{t=0} - G^\varkappa_t* \nab\crprd\EV^{\supN}|_{t=0}\big)(\sV) % \cr & 
+ c\varkappa^2\! \!\int_0^t\!\! \Big(\!G^\varkappa_{t-t'}* \HV^{\supN}(t',\,.\,)\Big)(\sV)\drm{t'} 
,\\
\hspace{-1truecm}
{\EV^{\supN}(t,\sV)} = %
& \label{eq:EsolKG}
\big(\tpddct G^\varkappa_t * \EV^{\supN}|_{t=0} + G^\varkappa_t* [\tpddct\EV^{\supN}]_{t=0}^{}\big)(\sV) % \cr&
 + c\varkappa^2\!\!\int_0^t\!\! \Big(\!G^\varkappa_{t-t'}* \DV^{\supN}(t',\,.\,)\Big)(\sV)\drm{t'}\!.
\end{alignat}

 All these solutions are understood not pointwise, but in the sense of distributions.
 In particular, with initial data allowed in this generality, singularities will typically propagate along the 
forward initial lightcones of the particles.

 To couple the MBLTP field equations with the equations for point charge motions, we want the fields 
$\BV^{\supN} (t,\Cdot)$ and $\EV^{\supN} (t,\Cdot)$ to be typically as regular as possible, given the particle
initial data for the velocities, $\vV_{\!n}^\sigma(0)$, satisfying $|\vV_{\!n}^\sigma(0)|< c$, and for the 
positions, $\qV_n^\sigma(0)$, which have to be pairwise distinct.
 Thus we restrict the field initial data to be of the form $(\BV^{\supN}, \DV^{\supN}, \EV^{\supN}, \dot\EV^{\supN})(0,\,.\,)$
$=(\BV_0, \DV_0, \EV_0, \dot\EV_0)(0,\,.\,)$ 
$+\sum_\sigma\sum_{j=1}^{N_\sigma}(\BV_j^\sigma, \DV_j^\sigma, \EV_j^\sigma, \dot\EV_j^\sigma)(0,\,.\,)$. 
 Here,
$(\BV_0, \DV_0, \EV_0, \dot\EV_0)(0,\,.\,)$ {\color{black}is the evaluation at 
$t=0$ of a smooth finite-energy, source-free solution to the MBLTP field equations, which is globally bounded by
$(\mathcal{B}_0, \mathcal{D}_0, \mathcal{E}_0, \dot{\mathcal{E}}_0)$, 
and has Lipschitz constants $L_{\BV_0}$ and $L_{\EV_0}$.
 Such solutions exist a plenty for the vacuum MBLTP field equations.}
 For $j\in\{1,...,N_\sigma\}$, the
$(\BV_j^\sigma, \DV_j^\sigma, \EV_j^\sigma, \dot\EV_j^\sigma)(0,\,.\,)$ is the ``co-moving electromagnetic field'' of a 
fictitious point charge whose world line coincides with the tangent world line of the $j$-th point charge at $t=0$.
 For these initial data, the general solution formulas can be brought into a more explicit format; see below.
 {\color{black}The pertinent field initial data are obtained by simply setting $t=0$ in these solution formulas.}

%%%%%%%%%%%%%%%%%%%%%%%%%%%%%%%%%%%%%%%%%%%%%%%%%%%%%%%%%%%%%%
%%%%%%%%%%%%%%%%%%%%%%%%%%%%%%%%%%%%%%%%%%%%%%%%%%%%%%%%%%%%%%
  \subsection{The equations of point charge motion}\label{sec:relPTmech}
%%%%%%%%%%%%%%%%%%%%%%%%%%%%%%%%%%%%%%%%%%%%%%%%%%%%%%%%%%%%%%
%%%%%%%%%%%%%%%%%%%%%%%%%%%%%%%%%%%%%%%%%%%%%%%%%%%%%%%%%%%%%%%%%%%%

 The particle initial data $\qV_n^\alpha(0)$ and $\vV_{\!n}^\alpha(0)$ are continued into the future $t>0$
by the following equations of motion.
 Each point particle's position and velocity vectors are related by the usual formula
\begin{equation}
\textstyle\Ddt \qV_n^\alpha(t)
= \label{eq:dotQisV}
\vV_{\!n}^\alpha(t).
\end{equation}
{\color{black}The particle velocity $\vV_{\!n}(t)$ and momentum $\pV_n(t)$ are related by
the familiar formula
\begin{equation}
\pV_n^\alpha(t) 
= \label{eq:ELPvTOp}
m_\alpha^{} \frac{{\vV_{\!n}^\alpha}(t)}{\sqrt{1 - \frac{1}{c^2}|{\vV_{\!n}^\alpha}(t)|^2}},
\end{equation}
see \cite{LorentzENCYCLOP}, \cite{einsteinA}, \cite{Poincare};
here, $m_\alpha^{}\neq 0$ is the \emph{(bare) inertial rest mass} of species $\alpha$.}
 The momentum $\pV^\alpha_n(t)$ in turn changes with time according to Newton's second law,
\begin{equation}
\textstyle\Ddt   \pV_n^\alpha(t) 
= \label{eq:EinsteinNewtonEQofMOT}
\fV_n^\alpha(t). 
\end{equation}
 The BLTP force $\fV_n^\alpha(t)$ on the $n$-th point charge of species $\alpha$ has been worked out in 
\cite{KTZonBLTP} and \cite{Kie19}, and is given by
\begin{alignat}{1}
\fV^\alpha_{n}(t)
 = \label{eq:fnalpha}
  & 
 - {\textstyle\Ddt} {\displaystyle\int_{B_{ct}(\qV_n^\alpha(0))}\hspace{-10pt} \!\!
\left( \PiV^{\mbox{\tiny{field}}}_{\alpha,n} (t,\sV) -
\PiV^{\mbox{\tiny{field}}}_{\alpha,n}(0,\sV-\olqV_n^\alpha(t))\right)\!\drm^3{s}} \cr
  & + e_\alpha^{} 
\Bigl[\EV_{0}(t,\qV_n^\alpha(t)) + \textstyle{\frac{1}{c}}\vV_{\!n}^\alpha(t) \crprd\BV_{0}(t,\qV_n^\alpha(t))\Bigr] 
\cr
   & + e_\alpha^{} \textstyle{\sum\limits_{\genfrac{}{}{0pt}{2}{1\leq j\leq N_\alpha }{j\neq n}}} 
\Bigl[\EV_{j}^\alpha(t,\qV_n^\alpha(t)) + \textstyle{\frac{1}{c}}\vV_{\!n}^\alpha(t) \crprd\BV_{j}^\alpha(t,\qV_n^\alpha(t))\Bigr] 
\cr
   & + e_\alpha^{} \textstyle{\sum\limits_{\beta \neq \alpha}  }
\textstyle{\sum\limits_{\ 1\leq j\leq N_\beta}} 
\Bigl[\EV_{j}^\beta(t,\qV_n^\alpha(t))+\textstyle{\frac{1}{c}}\vV_{\!n}^\alpha(t) \crprd\BV_{j}^\beta(t,\qV_n^\alpha(t))\Bigr],
\end{alignat}
where the fields $\EV_0(t,\sV)$ and $\BV_0(t,\sV)$ belong to the source-free solution of the MBLTP field 
equations, launched by the source-free initial data stipulated above, and which is given by 
\refeq{eq:HforN}--\refeq{eq:EsolKG} with all source terms set to zero, while the 
fields $\EV_j^\alpha(t,\sV)$ and $\BV_j^\alpha(t,\sV)$ (resp.\ 
$\EV_j^\beta(t,\sV)$ and $\BV_j^\beta(t,\sV)$) belong to the solution of the MBLTP field equations having the
$j$-th charge of the $\alpha$-th (resp.\ $\beta$-th) species as only source, 
launched by the pertinent sourced initial data stipulated above, and which will be written semi-explicitly
below; 
 finally, 
$\PiV^{\mbox{\tiny{field}}}_{\alpha,n} (t,\sV)$ is the \emph{electromagnetic field momentum density}
of the MBLTP fields 

\noindent
of the $n$-th charge of the $\alpha$-th species, with (omitting $n$ and $\alpha$)
\begin{equation}
\textstyle
c \PiV^{\mbox{\tiny{field}}}
= \label{eq:PiMBLTP}
\DV\crprd\BV + \EV\crprd\HV - \EV\crprd\BV 
- \varkappa^{-2} \big(\nab\cdot\EV\big)\Big(\nab\crprd\BV - \frac{1}{c}\pdt\EV\Big).
\end{equation}

 To evaluate the force formula \refeq{eq:fnalpha}, the following representation of the
 sourced solutions of the MBLTP field equations for $t> 0$ is helpful.

 A particle's $\Csp^{1,1}$ world line $t\mapsto \qV(t)$ for $t> 0$ will be extended to $t\leq 0$ by the auxiliary world line  
$t\mapsto \qV(0)+\vV(0)t \equiv \olqV(t)$.
 For a.e. $t\in\Rset$, let $\aV^{}(t):= \Ddt\vV^{}(t)$.
 Note that by Rademacher's theorem the $t$-derivative of a Lipschitz continuous map $t\mapsto\vV^{}(t)$ exists 
almost everywhere; for our partly actual, partly auxiliary map $t\mapsto\vV^{}(t)$ it typically does not exist at $t=0$.
Then:

The MBLTP field solutions $\BV_n^\alpha(t,\sV)$ and $\EV_n^\alpha(t,\sV)$ for $t\geq 0$ are given by (cf.\ \cite{GratusETal})
(again omitting the indices $\alpha$ and $n$) 
\begin{alignat}{1}
\hspace{-20pt}
{\EV(t,\sV)} & =\; \label{eq:EjsolMBLTP} 
 e^{}\varkappa^2\tfrac{1}{8\pi}\tfrac{\nV(\qV^{},\sV)-{\vV^{}}/{c}}{1-\nV(\qV^{},\sV)\cdot{\vV}/{c}}
\Big|_{\mathrm{ret}}
\, -  
e^{} \varkappa^2 \tfrac{1}{4\pi} \int_{-\infty}^{t^\mathrm{ret}(t,\sV)}
 c  \mathbf{K}_{{\qV}(t'),{\vV}(t')}(t',t,\sV)  \drm{t'} , \\ 
\hspace{-20pt}
{\BV(t,\sV)} & =\; \label{eq:BjsolMBLTP}
 e^{} \varkappa^2 \tfrac{1}{8\pi}
\tfrac{\color{black}{\vV^{}}\crprd\nV(\qV^{},\sV)/{c}}{1-\nV(\qV^{},\sV)\cdot{\vV}/{c}}
\Big|_{\mathrm{ret}}
\, -
e^{} \varkappa^2\tfrac{1}{4\pi} \int_{-\infty}^{t^\mathrm{ret}(t,\sV)}
{\vV^{}(t')}\crprd \mathbf{K}_{{\qV}(t'),{\vV}(t')}(t',t,\sV)
% NB: - \vV(t')\crprd\vV^{}(t')(t-t') = NULL
\drm{t'} ,
\end{alignat}
with the abbreviation 
\begin{alignat}{1}
\mathbf{K}_{\tilde{\qV},\tilde{\vV}}(\tilde{t},t,\sV)  := \label{eq:KbfTILDE}
\tfrac{J_2\!\bigl(\varkappa\sqrt{c^2(t-\tilde{t})^2-|\sV-\tilde{\qV}|^2 }\bigr)}{{c^2(t-\tilde{t})^2-|\sV-\tilde{\qV}|^2}^{\phantom{n}} }
 \left(\sV-\tilde{\qV}- (t-\tilde{t})\tilde{\vV}\right),
\end{alignat}
where $\nV(\qV,\sV) := \frac{\sV-\qV}{|\sV-\qV|}$ for $\qV\neq \sV$,
   % $\gamma^2 = 1/({1-\frac{|\vV^{}(t)|^2}{c^2}})$,
and where ``$|_{\mathrm{ret}}$'' means $(\qV,\vV,\aV)= (\qV,\vV,\aV)(t^{\mathrm{ret}})$ 
with $t^{\mathrm{ret}}(t,\sV)$ defined implicitly by $c(t-t^{\mathrm{ret}}) = |\sV-\qV(t^{\mathrm{ret}})|$;
note that $t^\mathrm{ret}(t,\sV)<t$.
Note also that $t'\mapsto \sqrt{c^2(t-t')^2-|\sV-\qV(t')|^2}$ is positive and monotone decreasing to zero
on $(-\infty,t^\mathrm{ret}(t,\sV)]$ and  asymptotic to
 $\sqrt{c^2-|\vV(0)|^2}|t'|$ as $t'\downarrow -\infty$.

 To evaluate the radiation-reaction force term, we also need 
the MBLTP field solutions $\DV^\alpha_n(t,\sV)$ and $\HV^\alpha_n(t,\sV)$ for $t> 0$, which are for a.e. $|\sV-\qV(t)|>0$
given by the Li\'enard--Wiechert formulas (see \cite{lienard}, \cite{Wiechert}) (omitting $\alpha$ and $n$)
\vskip-.6truecm
\begin{alignat}{1}
% \hskip-1truecm
{\DV(t,\sV)} &=\label{eq:LWsolD}
e^{}\frac{1}{4\pi}\frac{\Big[\!{1-\frac{|\vV^{}|^2}{c^2}}\!\Big]
\frac{{\nV(\qV^{},\sV)}_{\phantom{!\!}}
-{\vV^{}}/{c}}{ |\sV-\qV^{}|^2} +\frac{\nV(\qV^{},\sV)\crprd
\bigl[\bigl(\nV(\qV^{},\sV)_{\phantom{!\!}}-{\vV^{}/c}\bigr)\crprd\aV^{}\bigr]}{c^2|\sV-\qV^{}|}
}{\bigl(\textstyle{1-\nV(\qV^{},\sV)\cdot {\vV^{}}/{c}}\bigr)^{\!3}}
\Biggl.\Biggr|_{\mathrm{ret}}\hskip-1truecm
\\
\hskip-1truecm
{\HV(t,\sV)} 
&= \label{eq:LWsolH}
        \nV(\qV^{},\sV)|_{_{\mathrm{ret}}}\crprd {\DV(t,\sV)}
\, .\vspace{-10pt}
\end{alignat}
{\color{black}We remark that after an integration by parts of the Bessel kernel integrals, cf.\ \cite{GratusETal},
\refeq{eq:EjsolMBLTP}, \refeq{eq:BjsolMBLTP} reduce to the Lienard-Wiechert fields \refeq{eq:LWsolD}, \refeq{eq:LWsolH} 
when $\varkappa\to\infty$.}

 We also need integral representations for $\nab\cdot\EV^\alpha_n$ and $\nab\crprd\BV^\alpha_n - \frac{1}{c} \frac{\partial\;}{\partial t}\EV^\alpha_n$.
 Omitting $\alpha$ and $n$ again, we have 
\begin{eqnarray}
\hspace{-1truecm}
\qquad\qquad
\tfrac{4\pi}{e\varkappa^2}
 \nab\cdot\EV(t,\sV) \!\!\!&=&\!\!\!\; 
\label{eq:LWsolPHI} 
 \tfrac{1}{\bigl(1-\nV(\qV,\sV)\cdot {\vV^{}}/{c}\bigr)} \tfrac{1}{|\sV-\qV|} \Bigl.\Bigr|_{\mathrm{ret}} - 
 \varkappa \int_{-\infty}^{t^\mathrm{ret}(t,\sV)} \!\!\!\!
 \mathrm{K}_{\qV(t')}(t',t,\sV) c \drm{t'} ,
\\
\hspace{-1truecm} 
\tfrac{4\pi}{e\varkappa^2}
\big(\nab\crprd\BV - {\textstyle{\frac{1}{c}\pdt}}\EV\big)(t,\sV) \!\!\!&=&\!\!\!\;  
\label{eq:LWsolA}
 \tfrac{1}{\bigl(1-\nV(\qV,\sV)\cdot {\vV^{}}/{c}\bigr)}
\tfrac{1}{|\sV-\qV|}\tfrac{\vV^{}}{c} \Bigl.\Bigr|_{\mathrm{ret}} - 
 \varkappa  \int_{-\infty}^{t^\mathrm{ret}(t,\sV)} \!\!\!\!
\mathrm{K}_{\qV(t')}(t',t,\sV)   \vV^{}(t')\drm{t'} 
\end{eqnarray}
with the abbreviation 
\begin{alignat}{1}
\mathrm{K}_{\tilde{\qV}}(\tilde{t},t,\sV)  := \label{eq:KrmTILDE}
\tfrac{J_1\!\bigl(\varkappa\sqrt{c^2(t-\tilde{t})^2-|\sV-\tilde{\qV}|^2 }\bigr)}{\sqrt{{c^2(t-\tilde{t})^2-|\sV-\tilde{\qV}|^2}}^{\phantom{n}} }  .
\end{alignat}

 The representations \refeq{eq:EjsolMBLTP} and \refeq{eq:BjsolMBLTP} are explicit enough to allow the computation of 
the Lorentz force terms at r.h.s.\refeq{eq:fnalpha}. 
 To compute the ``self''-force terms at r.h.s.\refeq{eq:fnalpha} we note that with the
representations \refeq{eq:EjsolMBLTP}--\refeq{eq:LWsolA} the volume integrations at r.h.s.\refeq{eq:fnalpha}
can be reduced (essentially) to one-dimensional integrals over time by
switching to ``retarded spherical co-ordinates;'' see \cite{KTZonBLTP} for details. 
 This gives
\begin{alignat}{1}
\label{eq:PIminusPInullDECOMPexplicated}
&{16\pi^2 }{\displaystyle\int_{B_{ct}(\qV_n(0))}\!\!
 \left(\PiV^{\mbox{\tiny{field}}}_{\alpha,n}(t,\sV)-{\PiV}^{\mbox{\tiny{field}}}_{\alpha,n}(0,\sV-\olqV^\alpha_n(t))\right)\!\drm^3{s}}
=  \\ \notag
& \qquad\qquad\qquad
  {e_\alpha^2}
 \;{\textstyle\sum\limits_{k=0}^2}\; c^{2-k}
\displaystyle  \int_0^{t}\! 
\Bigl[{\mathbf{Z}}_{\boldsymbol{\xi}_n^\alpha}^{[k]}\big(t,t^{\mbox{\tiny{r}}}\big)
-\!
{\mathbf{Z}}_{\oli{{\boldsymbol{\xi}}}_n^\alpha}^{[k]}\big(t, t^{\mbox{\tiny{r}}}\big)\Bigr]
(t- t^{\mbox{\tiny{r}}})^{2-k} \drm{t^{\mbox{\tiny{r}}}} ,
\end{alignat}
where the kernels ${\mathbf{Z}}_{\boldsymbol{\xi}_n^\alpha}^{[k]}\big(t, t^{\mbox{\tiny{r}}}\big)$ are explained in Appendix A.
 The time derivative of \refeq{eq:PIminusPInullDECOMPexplicated} which appears at r.h.s.\refeq{eq:fnalpha} 
is now quite explicit --- note that there will be three contributions, one from the upper limit of integrations, and two from the integrand. 
 We remark that the  $t$-derivative of 
${\mathbf{Z}}_{\boldsymbol{\xi}_n^\alpha}^{[k]}\big(t, t^{\mbox{\tiny{r}}}\big)$ does not act on the functions 
$(\qV_n^\alpha, \vV_n^\alpha,\aV_n^\alpha)\equiv \boldsymbol{\xi}_n^\alpha$, which enter in 
${\mathbf{Z}}_{\boldsymbol{\xi}_n^\alpha}^{[k]}\big(t, t^{\mbox{\tiny{r}}}\big)$ only as functions of the integration variable $t^{\mathrm{r}}$.
 
%%%%%%%%%%%%%%%%%%%%%%%%%%%%%%%%%%%%%%%%%%%%%%%%%%%%%%%%%%%%%%
%%%%%%%%%%%%%%%%%%%%%%%%%%%%%%%%%%%%%%%%%%%%%%%%%%%%%%%%%%%%%%
  \subsection{The joint initial value problem is well-posed}\label{sec:wellposed}
%%%%%%%%%%%%%%%%%%%%%%%%%%%%%%%%%%%%%%%%%%%%%%%%%%%%%%%%%%%%%%
%%%%%%%%%%%%%%%%%%%%%%%%%%%%%%%%%%%%%%%%%%%%%%%%%%%%%%%%%%%%%%%%%%%%

 In \cite{KTZonBLTP} the joint initial value problem formulated in the previous two sub-subsections 
was shown to be locally well-posed. 
 An announcement is available in \cite{Kie19}.
 For the convenience of the reader, we briefly summarize the main ingredients.

 A key feature of the force term \refeq{eq:fnalpha} is that the acceleration $\aV_n^\alpha(t)$ only enters
through the Li\'enard--Wiechert formulas for the fields $\DV_n^\alpha$ and $\HV_n^\alpha$, and there
only in a \emph{linear} fashion. 
 Therefore, the acceleration $\aV_n^\alpha(t)$, or rather its history $t'\mapsto\aV_n(t')$ for $0<t'\leq t$, 
enters only through the radiation-reaction force term in the first line at r.h.s.\refeq{eq:fnalpha}, and in 
a linear manner.
 The Lorentz force terms only involve the (pertinent histories of the)
positions and velocities of the (at most) two different charged point particles involved.
 As a result, Newton's second law, together with \refeq{eq:ELPvTOp} and \refeq{eq:fnalpha}, becomes
a linear \emph{Volterra integral equation} for the acceleration $\aV_n^\alpha$ as a functional of the 
histories (from the initial time on until $t$) of the positions and velocities of all charged point particles.
 In \cite{KTZonBLTP} it is proved that a unique solution exists which depends Lipschitz-continuously on those
position and velocity histories. 

 This in turn results in the well-posedness of the joint initial value problem of BLTP electrodynamics.
 The evolution is global unless in a finite amount of time:

a) two or more point particles meet at the same location;

b) a point charge reaches the speed of light;

c) the acceleration of a point charge becomes infinite.

%%%%%%%%%%%%%%%%%%%%%%%%%%%%%%%%%%%%%%%%%%%%%%%%%%%%%%%%%%%%%%
%%%%%%%%%%%%%%%%%%%%%%%%%%%%%%%%%%%%%%%%%%%%%%%%%%%%%%%%%%%%%%
  \subsection{The main conservation laws}\label{sec:conservation}
%%%%%%%%%%%%%%%%%%%%%%%%%%%%%%%%%%%%%%%%%%%%%%%%%%%%%%%%%%%%%%
%%%%%%%%%%%%%%%%%%%%%%%%%%%%%%%%%%%%%%%%%%%%%%%%%%%%%%%%%%%%%%%%%%%%
 
 In BLTP electrodynamics,
the total energy density $ \veps(t,\sV):=\veps^{\mbox{\tiny{field}}}(t,\sV) + \veps^{\mbox{\tiny{points}}}(t,\sV)$,
the total momentum density 
$\PiV(t,\sV):=\PiV^{\mbox{\tiny{field}}}(t,\sV) + \PiV^{\mbox{\tiny{points}}}(t,\sV)$,
and the total stress field $T(t,\sV):=T^{\mbox{\tiny{field}}}(t,\sV)+T^{\mbox{\tiny{points}}}(t,\sV)$, 
jointly satisfy:

(i) the \emph{local conservation law for the total energy} 
\begin{equation}
\textstyle
\pdt \veps (t,\sV) + c^2 \nab\cdot \PiV(t,\sV) 
= \label{eq:ENconservation}
0,
\end{equation}

(ii) the \emph{local conservation law for the total momentum}, 
\begin{equation}
\textstyle
\pdt \PiV(t,\sV) + \nab\cdot T (t,\sV)
= \label{eq:MOMconservation}
\NullV,
\end{equation}

(iii) the \emph{local conservation law for total angular-momentum} 
\begin{equation}
\textstyle
\pdt \sV\crprd\PiV(t,\sV) 
+ \sV\crprd \nab\cdot  T (t,\sV)
= \label{eq:ANGMOMconservation}
\NullV.
\end{equation}

 The field momentum density is given in \refeq{eq:PiMBLTP}. 
 The \emph{field energy density} is given by
\begin{equation}
\veps^{\mbox{\tiny{field}}}
= \label{eq:TooMBLTP}
 \BV\cdot\HV +  \EV\cdot\DV  - \tfrac12\big(|\BV|^2 +|\EV|^2\big) - 
 \tfrac{1}{2\varkappa^2}  \Big[\big(\nab\cdot\EV\big)^2 + \big|\nab\crprd\BV - {\textstyle\frac1c\pdt}\EV\big|^2 \Big],
\end{equation}
and the \emph{symmetric stress field tensor of the electromagnetic fields} by
\begin{equation}\begin{array}{llll}
\hspace{-10pt}
 T^{\mbox{\tiny{field}}}
= \label{eq:TMBLTP}
&\!\!
 \tfrac12\big(\BV\cdot\HV + \HV\cdot\BV -|\BV|^2\big){\rm Id} -
 \BV\otimes\HV - \HV\otimes\BV + \BV\otimes\BV \; + \\ 
&\!\!
 \tfrac12\big(\DV\cdot\EV +\EV\cdot\DV -|\EV|^2 \big){\rm Id} -
 \EV\otimes\DV - \DV\otimes\EV + \EV\otimes\EV \; + \\
&\!\!
\tfrac{1}{\varkappa^2} \Big[
\tfrac12\big|\nab\crprd\BV - \frac{1}{c}\pdt\EV\big|^2 {\rm Id}
 - \Big(\nab\crprd\BV - \frac{1}{c}\pdt\EV\Big)\otimes\Big(\nab\crprd\BV- \frac{1}{c}\pdt\EV\Big)
                        \Big]\; - \qquad\\
&\!\!
\tfrac{1}{\varkappa^2}\frac12 \big(\nab\cdot\EV\big)^2 {\rm Id};
\end{array}
\end{equation}
note that we have defined $T^{\mbox{\tiny{field}}}$ with the opposite sign compared to the historical
      convention introduced by Maxwell to define the ``Maxwell stress tensor.''

 The  \emph{momentum vector-density of the point particles} is given by
\begin{equation}
\PiV^{\mbox{\tiny{points}}}
(t,\sV)
= \label{eq:TokCHARGES}
\textstyle \sum\limits_\alpha {\sum\limits_{\ 1\leq j\leq N_\alpha}} 
 \pV_{n}^\alpha(t)\; \delta_{\qV_n^\alpha(t)}(\sV),
\end{equation}
the \emph{energy density of the point particles} by 
\begin{equation}
 \veps^{\mbox{\tiny{points}}} (t,\sV)
= \label{eq:TooCHARGES}
\textstyle \sum\limits_\alpha  m_\alpha^{}c^2 {\sum\limits_{\ 1\leq j\leq N_\alpha}} 
{\sqrt{1 + \tfrac{|\pV_{n}^\alpha(t)|^2}{m_\alpha^{2}c^2}}}\; \delta_{\qV_n^\alpha(t)}(\sV),
\end{equation}
and the \emph{symmetric stress tensor of the point particles} is given by
\begin{equation}
T^{\mbox{\tiny{points}}} (t,\sV)
= \label{eq:TchargeSTRESS}
\textstyle \sum\limits_\alpha \frac{1}{m_\alpha^{}} {\sum\limits_{\ 1\leq j\leq N_\alpha}} 
\frac{\pV_{n}^\alpha(t)\otimes \pV_{n}^\alpha(t)}{\sqrt{1 + \frac{|\pV_{n}^\alpha(t)|^2}{m_\alpha^{2}c^2}}}\; \delta_{\qV_n^\alpha(t)}(\sV).
\end{equation}

 By hypothesis, the vacuum fields decay rapidly enough when $|\sV|\to\infty$ to guarantee a finite
field energy and momentum, and thus the total momentum and energy are conserved in BLTP electrodynamics.
 With sufficient decay of the fields at spatial infinity, also total angular-momentum is conserved.
 These global conservation laws follow from the local laws by integration w.r.t.\ d$^3s$.

 Finally, every point charge satisfies its own conservation law
\begin{equation}
\textstyle
     \forall\;\alpha,n:\quad   \pdt\delta_{\qV^\alpha_n(t)}(\sV) + \nab\cdot\left( \delta_{\qV^\alpha_n(t)}(\sV){\vV^\alpha_{\!\!n}}(t)\right)
=\label{eq:continuityOFmotion}
	0
\end{equation}
in the sense of distributions. 
 By the linearity of the continuity equation \refeq{eq:continuityOFmotion}, also
the distribution-valued fields 
$ e_\alpha {\sum_{n=1}^{N_\alpha}} \delta_{\qV^\alpha_n(t)}(\sV)\equiv \varrho_\alpha(t,\sV)$
and 
$ e_\alpha {\sum_{n=1}^{N_\alpha}} \delta_{\qV^\alpha_n(t)}(\sV){\vV^\alpha_{\!\!n}}(t)\equiv\jV_\alpha(t,\sV)$
jointly satisfy the continuity equation  
\begin{equation}
\textstyle
\forall\,\alpha:\quad        \pdt\varrho_\alpha (t,\sV) + \nab\cdot\jV_\alpha(t,\sV)
=\label{eq:MrhojLAWalpha}
	0. 
\end{equation}
 At last, defining the total electric point charge density $\sum_\alpha \varrho_\alpha (t,\sV)\equiv \varrho(t,\sV)$,  
and the total point charge current vector-density $\sum_\alpha \jV_\alpha(t,\sV) \equiv\jV(t,\sV)$,
they jointly satisfy 
\begin{equation}
\textstyle
        \pdt\varrho (t,\sV) + \nab\cdot\jV(t,\sV)
=\label{eq:MrhojLAW}
	0. \qquad \mbox{(\emph{Law\ of\ Charge\ Conservation})}
\end{equation}

 The conservation laws \refeq{eq:MrhojLAW} and \refeq{eq:MrhojLAWalpha} have a meaning also in the Vlasov continuum 
limit; the law \refeq{eq:continuityOFmotion} obviously not.

%%%%%%%%%%%%%%%%%%%%%%%%%%%%%%%%%%%%%%%%%%%%%%%%%%%%%%%%%%%%%%%%%%%%
%%%%%%%%%%%%%%%%%%%%%%%%%%%%%%%%%%%%%%%%%%%%%%%%%%%%%%%%%%%%%%%%%%%%
%%%%%%%%%%%%%%%%%%%%%%%%%%%%%%%%%%%%%%%%%%%%%%%%%%%%%%%%%%%%%%%%%%%%
	\section{The Vlasov approximation for $N\gg 1$}\label{sec:VLASOVapprox}\vspace{-5pt}
%%%%%%%%%%%%%%%%%%%%%%%%%%%%%%%%%%%%%%%%%%%%%%%%%%%%%%%%%%%%%%%%%%%%
%%%%%%%%%%%%%%%%%%%%%%%%%%%%%%%%%%%%%%%%%%%%%%%%%%%%%%%%%%%%%%%%%%%%
%%%%%%%%%%%%%%%%%%%%%%%%%%%%%%%%%%%%%%%%%%%%%%%%%%%%%%%%%%%%%%%%%%%%
%%%%%%%%%%%%%%%%%%%%%%%%%%%%%%%%%%%%%%%%%%%%%%%%%%%%%%%%%%%%%%%%%%%%
%%%%%%%%%%%%%%%%%%%%%%%%%%%%%%%%%%%%%%%%%%%%%%%%%%%%%%%%%%%%%%%%%%%%
%%%%%%%%%%%%%%%%%%%%%%%%%%%%%%%%%%%%%%%%%%%%%%%%%%%%%%%%%%%%%%%%%%%%
%	\section{The Vlasov limit $N\to\infty$}\label{sec:VLASOVlimit}\vspace{-5pt}
%%%%%%%%%%%%%%%%%%%%%%%%%%%%%%%%%%%%%%%%%%%%%%%%%%%%%%%%%%%%%%%%%%%%
%%%%%%%%%%%%%%%%%%%%%%%%%%%%%%%%%%%%%%%%%%%%%%%%%%%%%%%%%%%%%%%%%%%%
%%%%%%%%%%%%%%%%%%%%%%%%%%%%%%%%%%%%%%%%%%%%%%%%%%%%%%%%%%%%%%%%%%%%
%
 To study plasma, for which $N\gg 1$, it is neither feasible to solve for all the 
individual motions nor to follow them empirically. 
 Instead one switches the perspective and focusses on typical large-number type questions, such as:
``How many particles of species $\alpha$ reside in a small macroscopic
region about a space point $\sV$ at time $t$?'' ---  and:
``What is the coarse-grained statistics over their momenta?''.

 Answers to such large-number type questions are determined, theoretically, by integrating the counting
measures for the particle species over the relevant small macroscopic domains in $\sV$ and
$\pV$ space, respectively.
 The normalized counting measure of species $\alpha$ is the singular (with respect to Lebesgue measure)
empirical one-point measure on $\Rset^6$ given by
\begin{equation}
  \ul{\triangle}^{(1)}_{N_\alpha}(t,\sV,\pV)
 : =
  \frac{1}{N_\alpha} \textstyle\sum\limits_{k=1}^{N_\alpha}
   \delta_{\qV_k^{\alpha}(t)}(\sV) \delta_{\pV_k^{\alpha}(t)}(\pV) \,.
\label{empMEASalpha}
\end{equation}
 Every classical solution of $N$-point BLTP electrodynamics uniquely determines an empirical measure
\refeq{empMEASalpha} for all times $t$ for which the evolution lasts.

 When $N\gg 1$, then the normalized empirical measure $\ul{\triangle}_{N_\alpha}(t,\sV,\pV)$ may possibly 
be well-approximated by a smooth and gently varying normalized continuum density function $f_\alpha(t,\sV,\pV)$ satisfying 
a PDE in its indicated variables, more precisely: a Vlasov-type PDE system.
 This Vlasov PDE system in turn allows one to study the large scale structures and dynamics of plasma on time scales which
are not too long. 

 We now explain how the Vlasov--Maxwell--BLTP equations can be obtained in a limit as $N\to\infty$,
and what is needed to make this rigorous.
%===========================================================
\subsection{Evolution of empirical measures in BTLP electrodynamics}
 \label{secEMPmeasDYN}
%===========================================================

 The MBLTP field equations with point charge source terms, \refeq{eq:MdotD} and \refeq{eq:MdivD}, are already of 
the type \refeq{eq:rVMdotD} and \refeq{eq:rVMdivD}, yet
with $f_\alpha(t,\sV,\pV)$ replaced by $\ul{\triangle}^{(1)}_{N_\alpha}(t,\sV,\pV)$, i.e.
\begin{eqnarray}
  -\pdt{{\DV}^{\supN}(t,\sV)} + c\pds\times{\HV}^{\supN}(t,\sV) 
  \!\!&=&\!\!
    % 4 \pi
           {\textstyle\sum\limits_\alpha^{}} N_\alpha e_\alpha
     \int_{{\Rset}^3} \vV_\alpha(\pV) \ul{\triangle}^{(1)}_{N_\alpha} (t,\sV,\pV)\dd^3{p}\,,
  \label{eq:MdotDemp}
  \\
        \pds\cdot{\DV}^{\supN}(t,\sV)
  \!\!&=&\!\!
        %  4 \pi
        {\textstyle\sum\limits_\alpha^{}} N_\alpha e_\alpha \int_{{\Rset}^3}\ul{\triangle}^{(1)}_{N_\alpha} (t,\sV,\pV)\dd^3{p},
  \label{eq:MdivDemp}
\end{eqnarray}
 It is thus clear that whenever $\ul{\triangle}^{(1)}_{N_\alpha} (t,\sV,\pV) \approx f_\alpha(t,\sV,\pV)$ in
a suitable Kantorovich--Rubinstein distance, then also the electromagnetic MBLTP fields of the $N$-body system will be
approximately equal to the pertinent Vlasov--MBLTP fields, in a weak sense.

 The next step is to set up the pertinent PDE for the empirical one-point measures $\uli{\triangle}^{(1)}_{N_\alpha}(t,\sV,\pV)$.
 From the equations for the particle motions in $N$-body BLTP electrodynamics, one finds that
the empirical one-point measure \refeq{empMEASalpha} and the empirical two-point measure
\begin{equation}
\uli{\triangle}^{(2)}_{N_\alpha}(t,\sV,\pV,\tilde{t},\tilde{\qV},\tilde{\pV})
 = \label{eq:TWOptEMPmeasCOtangBUNDL}  
\frac{1}{N_\alpha(N_\alpha-1)} \ 
 \textstyle{\sum\!\!\!\!\sum\limits_{\hspace{-.3truecm}\genfrac{}{}{0pt}{2}{1\leq j\neq k \leq N_\alpha }{}}} 
\delta_{\pV_j^\alpha(t)}(\pV)\delta_{\qV_j^\alpha(t)}(\sV)
\delta_{\pV_k^\alpha(\tilde{t})}(\tilde{\pV})\delta_{\qV_k^\alpha(\tilde{t})}(\tilde{\qV}),
\end{equation}
which also is uniquely determined by any classical solution of $N$-point BLTP electrodynamics,
jointly satisfy, in the sense of distributions, the integro-partial-differential 
identity
\begin{alignat}{1}
& \hspace{-1truecm}  \partial^{}_t \uli{\triangle}^{(1)}_{N_\alpha}(t,\sV,\pV)
+ \vV_\alpha^{}(\pV)\cdot\partial^{}_{\sV} \uli{\triangle}^{(1)}_{N_\alpha}(t,\sV,\pV)
\cr
& \hspace{.5truecm} + e_\alpha
\left(\EV_0(t,\sV) +
\txtcinv
\vV_\alpha(\pV)\times \BV_0(t,\sV)\right)\cdot\pdp \uli{\triangle}^{(1)}_{N_\alpha}(t,\sV,\pV)
\cr
 & \hspace{.5truecm} + e_\alpha^{} \textstyle{\sum\limits_{\beta \neq \alpha} } 
\Bigl[\EV^{\supNb}(t,\sV)+\textstyle{\frac{1}{c}}\vV_\alpha(\pV) \crprd\BV^{\supNb}(t,\sV)\Bigr]\cdot\pdp 
\uli{\triangle}^{(1)}_{N_\alpha}(t,\sV,\pV)\cr
   &  \hspace{-1truecm}
+ e_\alpha^{2}  \varkappa^2\tfrac{1}{8\pi}(N_\alpha-1)
 \int\!\!\!\int
\Bigl[
\tfrac{\nV(\tilde{\qV},\sV)-{\tilde{\vV}_\alpha}/{c}}{1-\nV(\tilde{\qV},\sV)\cdot{\tilde{\vV}_\alpha}/{c}}
+ \tfrac1c\vV_\alpha\crprd
\tfrac{{\tilde{\vV}_\alpha}\crprd\nV(\tilde{\qV},\sV)/{c}}{1-\nV(\tilde{\qV},\sV)\cdot{\tilde{\vV}_\alpha}/{c}}
\Bigr] \cdot\pdp 
\uli{\triangle}^{(2)}_{N_\alpha}(t,\sV,\pV,{t^\mathrm{ret}(t,\sV)},\tilde{\qV},\tilde{\pV}) \dd^3\tilde{p}\dd^3\tilde{q} 
\cr
   &\hspace{-1truecm} 
- e_\alpha^{2}  \varkappa^2\tfrac{1}{4\pi} (N_\alpha-1)\!\!\! \int\limits_{-\infty}^{t^\mathrm{ret}(t,\sV)}\!\!\!\int\!\!\!\int
\Bigl[c \mathbf{K}_{\tilde{\qV},\tilde{\vV}_\alpha}(\tilde{t},t,\sV) 
+  \vV_\alpha\crprd ( \tilde{\vV}_\alpha \crprd \mathbf{K}_{\tilde{\qV},\tilde{\vV}_\alpha}(\tilde{t},t,\sV) )
\Bigr] \cdot\pdp 
\uli{\triangle}^{(2)}_{N_\alpha}(t,\sV,\pV,\tilde{t},\tilde{\qV},\tilde{\pV}) \dd^3\tilde{p}\dd^3\tilde{q} \dd\tilde{t}\cr
\label{eq:DELTA1nalphaEQ} 
  &\hspace{-10pt}= 
  \tfrac{1}{N_\alpha} \textstyle\sum\limits_{n=1}^{N_\alpha}
\Bigl({\textstyle\Ddt}  {\displaystyle\int_{B_{ct}(\qV_n^\alpha(0))}\hspace{-10pt} \!\!
\left( \PiV^{\mbox{\tiny{field}}}_{\alpha,n} (t,\tilde{\sV}) -
\PiV^{\mbox{\tiny{field}}}_{\alpha,n}(0,\tilde{\sV}-\olqV_n^\alpha(t))\right)\!\drm^3\tilde{s}}\! \Bigr)\!\cdot\pdp 
  \delta_{\pV_n^{\alpha}(t)}(\pV)  \delta_{\qV_n^{\alpha}(t)}(\sV).\quad
\end{alignat}
 Here, 
\begin{alignat}{2}
{\EV^{\supNb}(t,\sV)} =& \label{eq:EsolKGemp}
N_\beta e_\beta \varkappa^2\tfrac{1}{8\pi} 
 \int\!\!\!\int
\tfrac{\nV(\tilde{\qV},\sV)-{\tilde{\vV}_\beta}/{c}}{1-\nV(\tilde{\qV},\sV)\cdot{\tilde{\vV}_\beta}/{c}}
\;\uli{\triangle}^{(1)}_{N_\beta}({t^\mathrm{ret}(t,\sV)},\tilde{\qV},\tilde{\pV}) \dd^3\tilde{p}\dd^3\tilde{q} 
\cr &
\quad
- N_\beta e_\beta \varkappa^2\tfrac{1}{4\pi}  \!\!\! \int\limits_{-\infty}^{t^\mathrm{ret}(t,\sV)}\!\!\!\int\!\!\!\int
 c\mathbf{K}_{\tilde{\qV},\tilde{\vV}_\beta}(\tilde{t},t,\sV) 
\uli{\triangle}^{(1)}_{N_\beta}(\tilde{t},\tilde{\qV},\tilde{\pV}) \dd^3\tilde{p}\dd^3\tilde{q} \dd\tilde{t}, \\
{\BV^{\supNb}(t,\sV)} = & \label{eq:BsolKGemp} 
N_\beta e_\beta \varkappa^2\tfrac{1}{8\pi} 
 \int\!\!\!\int
 %\tfrac1c\vV_\alpha\crprd
\tfrac{{\tilde{\vV}_\beta}\crprd\nV(\tilde{\qV},\sV)/{c}}{1-\nV(\tilde{\qV},\sV)\cdot{\tilde{\vV}_\beta}/{c}}
\uli{\triangle}^{(1)}_{N_\beta}({t^\mathrm{ret}(t,\sV)},\tilde{\qV},\tilde{\pV}) \dd^3\tilde{p}\dd^3\tilde{q} \cr
&
\quad
- N_\beta e_\beta \varkappa^2\tfrac{1}{4\pi}  \!\!\! \int\limits_{-\infty}^{t^\mathrm{ret}(t,\sV)}\!\!\!\int\!\!\!\int
  \tilde{\vV}_\beta\crprd \mathbf{K}_{\tilde{\qV},\tilde{\vV}_\beta}(\tilde{t},t,\sV) 
\uli{\triangle}^{(1)}_{N_\beta}(\tilde{t},\tilde{\qV},\tilde{\pV}) \dd^3\tilde{p}\dd^3\tilde{q} \dd\tilde{t}.
\end{alignat}
 Also the radiation-reaction term at r.h.s.\refeq{eq:DELTA1nalphaEQ} can be written explicitly as a nonlinear functional
$\cR\big[\uli{\triangle}^{(1)}_{N_\alpha}\big](t,\sV,\pV)$, but to do so will fill several pages (see Appendix A) and is 
not necessary for the argument we are going to make here.

%===========================================================
\subsection{{\color{black}Continuum and} Propagation-of-Chaos {\color{black}approximations}}
 \label{secVLASOVapprox}
%===========================================================

 Since both the empirical one-point and the empirical two-point measures are involved,
the system of equations \refeq{eq:DELTA1nalphaEQ}, indexed by $\alpha$, 
is not a closed system for the empirical one-point measures.
      However, {\color{black} if the empirical one-point measures can be well-approximated by smooth
density functions $f_\alpha$ --- which requires $N_\alpha\gg 1$ ---},
then the system of equations \refeq{eq:DELTA1nalphaEQ} is approximately equal to a \emph{closed, nonlinear} 
system of evolution equation for these one-point measures, as suggested by the following:\vspace{-5pt}
{\color{black}\begin{obs}\label{obs:empTOprod}
       Suppose $\uli{\triangle}^{(1)}_{N}(t,\pV^{},\qV^{}) \stackrel{N\to\infty}{\longrightarrow} f(t,\pV^{},\qV^{})$  
in the bounded Lipschitz distance for one-point measures.
       Then
$\uli{\triangle}^{(2)}_{N}(t,\pV^{},\qV^{},\tilde{t},\tilde\pV,\tilde\qV)
\stackrel{N\to\infty}{\longrightarrow} {f}(t,\pV^{},\qV^{}) {f}(\tilde{t},\tilde\pV,\tilde\qV)$ 
in the bounded Lipschitz distance for two-point measures.\vspace{-3pt}
\end{obs}
\begin{rem} 
Appendix B gives a precise formulation of this observation and its proof.
\end{rem}

 Of course, in BLTP electrodynamics 
the Lorentz forces from one point charge onto another are only bounded but not Lipschitz continuous (they have
a jump discontinuity when two point charges meet at the same point), yet this singularity is much less severe than the 
one of the divergent Coulomb forces in a non-relativistic plasma.
 Thus, at least for a large class of physically interesting scenarios (more on this below), we expect that even} for finite $N\gg 1$, 
if the error is initially small when we
approximate $\uli{\triangle}^{(1)}_{N_\alpha}(0,\pV^{},\qV^{})$ in \refeq{eq:DELTA1nalphaEQ}
and in \refeq{eq:MdotDemp}, \refeq{eq:MdivDemp}, and in \refeq{eq:EsolKGemp}, \refeq{eq:BsolKGemp}, with $f_\alpha(0,\pV^{},\qV^{})$,
and $\uli{\triangle}^{(2)}_{N_\alpha}(0,\pV^{},\qV^{},\tilde{t},\tilde\pV,\tilde\qV)$ with 
${f_\alpha}(0,\pV^{},\qV^{}) {f_\alpha}(\tilde{t},\tilde\pV,\tilde\qV)$ in the fourth and fifth line at l.h.s.\refeq{eq:DELTA1nalphaEQ},
{\color{black}then \emph{for a certain amount of time the error of the pertinent time-evolved approximation also remains small}.
 The evolution of the empirical one-point measures should then be well-approximated by a closed
system of equations for the $f_\alpha(t,\pV^{},\qV^{})$, registered in:}

\begin{obs}\label{obs:propagationOFchaos} 
 Replacement of $\uli{\triangle}^{(1)}_{N_\alpha}(t,\pV^{},\qV^{})$ in \refeq{eq:DELTA1nalphaEQ}
and in \refeq{eq:MdotDemp}, \refeq{eq:MdivDemp}, and in \refeq{eq:EsolKGemp}, \refeq{eq:BsolKGemp}, with $f_\alpha(t,\pV^{},\qV^{})$,
and $\uli{\triangle}^{(2)}_{N_\alpha}(t,\pV^{},\qV^{},\tilde{t},\tilde\pV,\tilde\qV)$ with 
${f_\alpha}(t,\pV^{},\qV^{}) {f_\alpha}(\tilde{t},\tilde\pV,\tilde\qV)$ in the fourth and fifth line at l.h.s.\refeq{eq:DELTA1nalphaEQ},
yields the \emph{closed, nonlinear} system of equations
\begin{alignat}{1}\label{VlasovNrad}
&\hspace{-1truecm} \left[ \partial^{}_t 
+ \vV_\alpha^{}(\pV)\cdot\partial^{}_{\sV} 
 + e_\alpha \left(\EV(t,\sV) + \txtcinv \vV_\alpha(\pV)\times \BV(t,\sV)\right)\cdot\pdp\right] f_\alpha(t,\sV,\pV)
 = \\ \notag
& \hspace{1.6truecm}
 e_\alpha \tfrac{1}{N_\alpha}
\left({\EV}^\alpha(t,\sV) + \txtcinv \vV_\alpha(\pV)\times {\BV}^\alpha(t,\sV)\right)\cdot\pdp  f_\alpha(t,\sV,\pV)
+ \cR\left[f_\alpha\right](t,\sV,\pV), 
\hspace{-10pt}
\end{alignat}
with the fields $\BV$ and $\EV$ solving the Vlasov--MBLTP \emph{field} equations 
{\color{black}\refeq{eq:rVMrotE}}--\refeq{eq:rVMdivD} \&\ \refeq{eq:BLTPvaclaw}, while the fields 
${\BV}^\alpha$ and ${\EV}^\alpha$ solve 
{\color{black}\refeq{eq:rVMrotE}}--\refeq{eq:rVMdivD} \&\ \refeq{eq:BLTPvaclaw} with $e_\sigma^{}=0$ for all $\sigma\neq\alpha$.
\end{obs}

\newpage

{\color{black}
\begin{rem}
Not all empirical one-point measures which initially are well approximated by a smooth continuum function
will remain well so-approximated.
 For instance, take a system of only electrons, i.e.\ all particles have the same charge. 
 Then an initial state in which all particles are distributed uniformly inside a ball of radius $R$ in physical and of radius
$\epsilon$ in momentum space (a product of two such distributions, say), with
electrostatic field initial data, will not remain static but lead to an explosion, sending the particles flying off radially 
in all directions. 
 The time-evolved empirical one-point measure, even if initially well approximated by the characteristic function of the balls
of radius $R$ and $\epsilon$ will very quickly become not well approximated by a continuous density 
function because of the rapid attenuation of the empirical density.
 An exception might be the electron plasma confined in a grounded cylinder by a strong uniform magnetic field, but in this paper we do not
consider plasma with boundary.
\end{rem}}

 Since a continuum approximation of $\uli{\triangle}^{(1)}_{N_\alpha}(t,\pV^{},\qV^{})$ with $f_\alpha(t,\pV^{},\qV^{})$
is sensible only for $N\gg 1$ and, {\color{black}by the last remark, only for essentially neutral plasma (if in unbounded space)},
and since in the replacement 
of $\uli{\triangle}^{(2)}_{N_\alpha}(t,\pV^{},\qV^{},\tilde{t},\tilde\pV,\tilde\qV)$ with 
${f_\alpha}(t,\pV^{},\qV^{}) {f_\alpha}(\tilde{t},\tilde\pV,\tilde\qV)$ one already neglects terms of $\cO(1/N_\alpha)$, 
one's first impulse might be to also neglect the Lorentz force with factor $1/N_\alpha$ at r.h.s.\refeq{VlasovNrad}, which yields
\begin{alignat}{1}\label{VLASOVrad}
\hspace{-1truecm}
\left[ \partial^{}_t 
+ \vV_\alpha^{}(\pV)\cdot\partial^{}_{\sV} 
+ e_\alpha
\left(\EV(t,\sV) +
\txtcinv
\vV_\alpha(\pV)\times \BV(t,\sV)\right)\cdot\pdp\right] f_\alpha(t,\sV,\pV)
 = \cR\left[f_\alpha\right](t,\sV,\pV),\hspace{-15pt}
\end{alignat}
with the fields solving the Vlasov--MBLTP \emph{field} equations {\color{black}\refeq{eq:rVMrotE}}--\refeq{eq:rVMdivD} \&\ \refeq{eq:BLTPvaclaw}. 
 This might indeed be a good approximation whenever the fields $\EV$ and $\BV$ are as large as the fields
$\EV^\alpha$ and $\BV^\alpha$, which scale like $\cO(N)$.
 However, in an overall neutral plasma there are many cancellations, and situations are in principle conceivable in
which $\EV$ and $\BV$ are of the same $\cO(1)$ as $\frac{1}{N_\alpha}\EV^\alpha$ and $\frac{1}{N_\alpha}\BV^\alpha$, 
in which case all Lorentz force terms in \refeq{VlasovNrad} are equally important.
 In situations in which $\EV$ and $\BV$ are essentially zero, they may even be negligible versus the 
$\frac{1}{N_\alpha}\EV^\alpha$ and $\frac{1}{N_\alpha}\BV^\alpha$ terms at r.h.s.\refeq{VLASOVrad};
see the appendix of \cite{MikiShadi} for a related discussion.
 In this last situation, the $1/N_\alpha$ error term for the replacement of $\Delta^{(2)}$ by 
the product of two $f_\alpha$ may possibly play a significant role.

 Supposing that the fields $\EV$ and $\BV$ are $\propto N$ large so that the  $\alpha$-th Lorentz force at r.h.s.\refeq{VLASOVrad} can
be neglected, we need one more approximation to arrive at the Vlasov--MBLTP equations. 
 Fortunately, from \refeq{eq:PIminusPInullDECOMPexplicated} it follows that the radiation-reaction force is initially zero and 
builds up only in the course of time; see also \cite{Kie19} and \cite{KTZonBLTP}.
 Thus for sufficiently short time scales  the term $\cR\left[f_\alpha\right](t,\sV,\pV)$ can be neglected, and 
\refeq{VLASOVrad} becomes \refeq{eq:rVMfEQs}.

 We have obtained the complete system of Vlasov--MBLTP equations --- 
by very plausible approximations to a well-defined dynamical $N$-body plus field theory.

%===========================================================
\section{On the asymptotic exactness of Vlasov theory}
 \label{secVlimit}
%===========================================================

 While our plausibility derivation of the Vlasov--MBLTP equations may seem compelling, for we did not sweep any infinities
or otherwise undefined and not well-definable expressions under the rug, there is still the possibility that subtler 
reasons will invalidate it. 
 In particular, even if the errors of the Vlasov approximation to the actual microscopic $N$-body plus field 
BLTP electrodynamics are initially small, our arguments do not convey how long the errors would remain small. 
 These concerns can only be relieved of by a rigorous control of the errors made by the Vlasov approximation
to the BLTP electrodynamics of $N$ charges and their electromagnetic fields.

 A related but less ambitious first step toward such a control could be a proof of the asymptotic exactness of the 
Vlasov--MBLTP equations as $N\to\infty$.
 Asymptotic exactness of a kinetic theory, if it can be proven to hold, always refers to the vanishing, when $N\to\infty$,
of the error made by replacing the $N$-body formulation with the kinetic model. 
 It does not require a more quantitative estimate of the error when $N$ is finite. 

 There is a cheap way of (formally) establishing asymptotic exactness of the Vlasov--MBLTP system of equations by
making the charges $N$-dependent, viz. $e_\alpha^{} = \tilde{e_\alpha}/\surd{N}$. 
 We then write ${\BV^{\supNb}(t,\sV)} = \sqrt{N}\, {\widetilde{\BV}^{\supNb}(t,\sV)}$ and
${\EV^{\supNb}(t,\sV)} = \sqrt{N}\, {\widetilde{\EV}^{\supNb}(t,\sV)}$, and 
also scale up both $\EV_0\propto \surd{N}$ and $\BV_0\propto \surd{N}$ (as per their initial data) to
include a vacuum electromagnetic field of comparable strength.
 With these replacements in \refeq{eq:DELTA1nalphaEQ}--\refeq{eq:BsolKGemp}, letting $N\to\infty$ such that 
$N_\alpha/N = \nu_\alpha^{} +\cO(1/N)$ for fixed $\nu_\alpha^{}\in(0,1)$, satisfying $\sum_\alpha \nu_\alpha^{} =1$, 
one formally obtains the Vlasov--MBLTP system of equations (with $\widetilde{\phantom{v}}$ atop the field and charge symbols, 
which can then be dropped).
 The radiation-reaction term at r.h.s.\refeq{eq:DELTA1nalphaEQ} vanishes in this limit $N\to\infty$. 

 There are two comments to be made about this procedure. 

 First, it is still a long way to go to make even this simple argument rigorous, for one needs to establish that the 
$N$-body BLTP electrodynamics evolves for sufficiently long times, at least typically, for instance by showing that the
set of initial data which lead to a singularity in finite time is of measure zero. 
 Presumably this is a hard problem.
 It is not completely solved yet even for the Newtonian $N$-body problem with Newtonian gravitational, or with
Coulombian electrical forces between the point particles. 

 Second, the rescaling of the charges is a mathematical convenience, but unless 
it can be shown to be equivalent to an (at least in principle) physically realizable procedure it does not
explain why a plasma, in which charges, masses, the speed of light, and other such parameters
are fixed, may be accurately described by the Vlasov--MBLTP equations. 
 We are not saying anything new here ---
 Balescu \cite{BalescuBOOK1} said similar things about such ``derivations'' of the Vlasov--Maxwell equations; cf.
\cite{RostokerRosenbluth60}.

 So the question is whether the mathematical rescaling of the charges and fields stated above is equivalent to a
physically realizable sequence of BLTP $N$-body systems on adjusted time and space and field scales, at least in principle.
 Unfortunately, there seems to be \emph{no} physically realizable  scaling with $N$ for which 
the Vlasov--MBLTP system is asymptotically exact as $N\to\infty$ in the BLTP $N$-body plus field electrodynamics.
 
 Here is the argument.
  Given the particle species which compose a plasma, only the number of particles in a species, $N_\alpha$, is at one's 
disposal, and how one populates the plasma with these particles.
 The species parameters $m_\alpha$ and $e_\alpha$ are fixed, the speed of light $c$ is fixed, too, and in BLTP electrodynamics 
also the Bopp parameter $\varkappa$ is a ``constant of nature'' and thus fixed; see \cite{CKP} for empirical lower bounds on its size.
 (In the next section we will inquire into the limit $\varkappa\to\infty$ of the Vlasov--MBLTP system, but that's a 
different issue.)
 Of course, as before we set $N_\alpha/N = \nu_\alpha^{} +\cO(1/N)$ for fixed $\nu_\alpha^{}\in(0,1)$, 
satisfying $\sum_\alpha \nu_\alpha^{} =1$.
 If one now asks how to adjust the space and time scales, and scales of the fields, with $N$, at least in principle, 
so as to eliminate $N$ from the Vlasov--MBLTP system, one realizes that 
the velocities need to be left unscaled with $N$ because the speed limit $c$ is.
 This means that space and time have to be scaled in the same way. 
 But then the BLTP law of the electromagnetic vacuum, with $\varkappa$ unscaled, implies that space and time scales are
not rescaled with $N$ at all.
 This leaves the rescaling of the electromagnetic fields. 
 The inhomogeneous Maxwell equations suggest that all the fields scale $\propto N$, but inspection of 
\refeq{eq:DELTA1nalphaEQ} now reveals that this factor $N$ will only feature in the Lorentz force term 
and therefore does not factor out of this transport equation. 

 The conclusion of this discussion is that the Vlasov--MBLTP system is presumably only asymptotically exact in the
sense of the mathematical rescaling of charges and fields, but not in a physically (at least in principle) realizable sense. 
 To rigorously assess its validity will require good finite-$N$ bounds on the propagation of the errors made when 
approximating the empirical measures with the continuous Vlasov densities. \vspace{-10pt}

%%%%%%%%%%%%%%%%%%%%%%%%%%%%%%%%%%%%%%%%%%%%%%%%%%%%%%%%%%%%%%%%%%%%
%%%%%%%%%%%%%%%%%%%%%%%%%%%%%%%%%%%%%%%%%%%%%%%%%%%%%%%%%%%%%%%%%%%%
%%%%%%%%%%%%%%%%%%%%%%%%%%%%%%%%%%%%%%%%%%%%%%%%%%%%%%%%%%%%%%%%%%%%
	\section{On the limit $\varkappa\to\infty$ of the Vlasov--Maxwell--BLTP system}\label{sec:kappaTOinfty}\vspace{-5pt}
%%%%%%%%%%%%%%%%%%%%%%%%%%%%%%%%%%%%%%%%%%%%%%%%%%%%%%%%%%%%%%%%%%%%
%%%%%%%%%%%%%%%%%%%%%%%%%%%%%%%%%%%%%%%%%%%%%%%%%%%%%%%%%%%%%%%%%%%%
%%%%%%%%%%%%%%%%%%%%%%%%%%%%%%%%%%%%%%%%%%%%%%%%%%%%%%%%%%%%%%%%%%%%

 Having argued, convincingly as we hope, that the Vlasov--Maxwell--BLTP system 
{\color{black}\refeq{eq:rVMfEQs}--\refeq{eq:rVMdivD} \&\ \refeq{eq:BLTPvaclaw}} is presumably asymptotically 
exact in a limit $N\to\infty$ of $N$-body BLTP electrodynamics at least in the mathematical sense of rescaled 
charges and fields, though not in a physically realizable sense, our next observation is that the
Vlasov--MBLTP system formally reduces to the Vlasov--Maxwell system when $\varkappa\to\infty$.
 Of course, \emph{supposing} BLTP electrodynamics is the correct classical theory of electromagnetism (which 
may well not be the case!), then $\varkappa$ is fixed by nature, and the issue is to find good bounds on
the discrepancies of the solutions produced by Vlasov--Maxwell versus Vlasov--Maxwell--BLTP theory with finite $\varkappa$. 
 Similar to the discussion about the error terms of the Vlasov approximation to the finite-$N$ BLTP electrodynamics, 
a simpler preliminary step to accomplish the desired goal is to show rigorously that Vlasov--Maxwell--BLTP theory reduces
to Vlasov--Maxwell theory when $\varkappa\to \infty$.
 Since this is a singular limit, it can only be accomplished for a restricted class of field initial data.

%%%%%%%%%%%%%%%%%%%%%%%%%%%%%%%%%%%%%%%%%%%%%%%%%%%%%%%%%%%%%%%%%%%%
%%%%%%%%%%%%%%%%%%%%%%%%%%%%%%%%%%%%%%%%%%%%%%%%%%%%%%%%%%%%%%%%%%%%
	\subsection{Compatibility constraints on the field data}\label{sec:physics} 
%%%%%%%%%%%%%%%%%%%%%%%%%%%%%%%%%%%%%%%%%%%%%%%%%%%%%%%%%%%%%%%%%%%%
%%%%%%%%%%%%%%%%%%%%%%%%%%%%%%%%%%%%%%%%%%%%%%%%%%%%%%%%%%%%%%%%%%%%

\noindent
 The Vlasov--Maxwell--BLTP system is of higher order than the Vlasov--Maxwell system, i.e.\ it requires 
field initial data $\BV(0,\,.\,)$ and $\DV(0,\,.\,)$ (constrained by (\ref{eq:rVMdivB}) and (\ref{eq:rVMdivD}))
as well as  $\EV_0(0,\,.\,)$ and $\dot\EV_0(0,\,.\,)$, while the Vlasov--Maxwell system does not have any freedom 
to choose the latter two fields once the fields $\BV(0,\,.\,)$ and $\DV(0,\,.\,)$ are prescribed.
 Thus for the Vlasov--MBLTP field equations we need a rule which expresses the initial data $\EV(0,\sV)$ and $\dot\EV(0,\sV)$ 
in terms of $\BV(0,\sV)$ and $\DV(0,\sV)$ in a manner compatible with the Vlasov--Maxwell theory when $\varkappa\to\infty$;
otherwise the limit won't exist.
 Explicitly, when $\varkappa\to\infty$, 
the map from $\BV(0,\sV)$ and $\DV(0,\sV)$ to $\EV(0,\sV)$ and $\dot\EV(0,\sV)$ must reduce to the Maxwell law 
$\EV(0,\sV) = \DV(0,\sV)$, and to the Maxwell--Lorentz equation 
\begin{eqnarray}
\qquad  (\pdt{\EV})(0,\sV) 
  \!\!&=&\!\!
    % 4 \pi
 c\pds\times{\BV}(0,\sV) 
-           {\textstyle\sum\limits_\alpha^{}} N_\alpha e_\alpha
     \int_{{\Rset}^3} \vV_\alpha(\pV) f_\alpha (0,\sV,\pV)\dd^3{p}\,;
  \label{eq:rVMdotDtNULL}
\end{eqnarray}
here, $N_\alpha$ may be replaced by $\nu_\alpha^{}$ if the limit $N\to\infty$ has been invoked with 
mathematical rescaling of charges and fields, as explained in the section on asymptotic exactness. 

%===========================================================
\section{Conclusions}\vspace{-5pt}
 \label{secConclusion}
%===========================================================

 In the previous sections, we have proposed a road map for the microscopic foundations of
the kinetic theory of special-relativistic plasma, and to emphasize that a rigorous derivation seems feasible.
 While the standard plasma physics literature is regrettably blas\'e about this issue, usually claiming the 
mathematically ill-defined symbolic system of Lorentz equations for $N$ point charges and their Maxwell--Lorentz fields as 
the microscopic foundations of the Vlasov--Maxwell system of equations,\footnote{Of historical interest, perhaps, is that Vlasov 
himself considered the relativistic Vlasov theory as more fundamental \cite{VlasovBOOK}.
  Needless to say that his view never caught on.\vspace{-5pt}}
we have argued that the Vlasov--Maxwell system is rather an approximation to the Vlasov--Maxwell--BLTP system when Bopp's 
$\varkappa \gg 1$, which in turn is an accurate kinetic continuum approximation to the microscopic $N$-body electrodynamics 
in a  BLTP vacuum when $N\gg 1$ and when the time scales are not too long.
 Moreover, we presented a radiation-reaction-corrected system of kinetic equations for the BLTP electrodynamics 
which, when $N\gg 1$, should be more accurate than the Vlasov--MBLTP system on a slightly larger than the 
usual Vlasov time scale, but still shorter than the time scale on which discrete particle effects kick in.
 To establish the degree of accuracy of this approximation is an ambitious program for future research, but we 
are confident that this can be carried out with mathematical rigor.

 {\color{black}The road to the validity of relativistic Vlasov theory which we propose to travel is via the control of 
the actual empirical one- and two-point measures of the plasma, by establishing the asymptotic
(as $N\to\infty$) exactness of their continuum approximation on time scales of interest for a suitable class of initial data; 
cf.\ \cite{Kie14}.
 The factorization of the two-point measures into products of one-point functions (``propagation of chaos'') in the 
continuum limit should in fact follow from the continuum limit for the one-point measures; cf.\ Appendix B for 
Lipschitz continuous forces.

 If the BLTP interactions would not have a jump discontinuity when two particles meet at the same location, 
but would instead be Lipschitz continuous, our approach would simplify because then one could directly invoke Lemma B.1.
 And if the electromagnetic forces from one particle onto another would not only be Lipschitz continuous but
actually vanish when the two particles meet at the same location, then one could work exclusively with the 
actual empirical one-point measures, as did Neunzert, and no propagation of chaos estimate would ever be needed.
 In any event the technical estimates would be greatly facilitated.
 The whole approach would then become a direct variation on the theme of the works \cite{EKR} and \cite{Golse}, which
have adapted the strategy pioneered by Neunzert \cite{NeunzertA}; cf.\ \cite{BraunHepp,Dobru,NeunzertB,SpohnBOOKa}.

 Frequently one finds claims, e.g. \cite{Ichimaru, HazeltineWaelbroeck}, that the smooth 
Vlasov approximation to these empirical densities would represent an {ensemble average}, and that the
(normalized) $f(t,\sV,\pV)$ is a probability density.
 (Incidentally, this is often called ``\emph{mean-field} approximation.'')
 However, since a single many-body system cannot interact with the ensemble of independent copies of itself,
to which it belongs, such a ``mean-field approximation'' can only become meaningful through a law of large numbers which 
guarantees that almost all members of the ensemble behave precisely like their ensemble mean in the infinitely many particles limit.
 Such a proof would involve working with an infinite BBGKY hierarchy and, also, proving propagation of chaos for the marginal
$n$-point probability measures.

 In our approach, no ensemble averaging, in particular not the BBGKY hierarchy of \emph{all the $n$-point functions} when $N\to\infty$,
is ever invoked.
 By avoiding the BBGKY hierarchy, the approach proposed here is conceptually more direct, and promises to be technically
simpler.}

   The accuracy of the approximation of the $N$-body BLTP electrodynamics by the kinetic Vlasov--Maxwell or 
Vlasov--Maxwell--BLTP model cannot be expected to last for arbitrarily long times, for initially small errors 
in the approximation tend to grow exponentially over time.
 On longer time scales, various deviations of the Vlasov continuum evolution, and also of the
radiation-reaction-corrected Vlasov evolution, from the empirical evolution will become visible. 
  It is expected that additional dissipative corrections to kinetic equations of the Balescu--Lenard--Guernsey type
(which frequently are approximated by a Landau type equation) are needed to capture this long-time regime.
  Their microscopic foundation is an even more ambitious program, as indicated by Spohn \cite{SpohnBOOKa}.
  A fresh recent attack on this front was made by Lancellotti \cite{Lancellotti09}; see also \cite{VelazquezWinter}.

 Although we have not supplied any rigorous arguments in this paper, we believe that the program described in this work
can be carried out in full rigor. 
 In particular, the well-posedness of BLTP electrodynamics with $N$ point charges has already been proved rigorously
in \cite{KTZonBLTP}, see also \cite{Kie19}, and we expect that the error control of the Vlasov approximation will be
accomplished in due time.
 This will mean a major advance in the rigorous microscopic foundations of relativistic kinetic theory
of plasma without breaking the Lorentz covariance at any point in the analysis.
 We recall that the state of affairs so far has been based on a regularization of the Lorentz electrodynamics with $N$ point charges, 
essentially the semi-relativistic Abraham--Lorentz model, which leads to a non-Lorentz covariant regularization of
the Vlasov--Maxwell equations by Golse \cite{Golse}. 
 A precursor to this work is our paper with Ricci \cite{EKR}, where we derived a scalar
caricature of special-relativistic Vlasov theory from a classical microscopic model of $N$ finite-size particles which 
interact with a scalar wave field.

 We also expect that the BLTP law of the electromagnetic vacuum can be replaced by the Born--Infeld law \cite{BornInfeldBb}, 
but its nonlinear character makes progress very slow. 
 So far only the static problem of $N$ point charges has been settled \cite{KieMBIinCMP} (see also \cite{BonheureETal}), 
while the dynamical analog \cite{Kie04, Kie12} is still waiting to be treated rigorously.
\smallskip

\noindent
\textbf{ACKNOWLEDGEMENT:} The authors thank Holly Carley, Markus Kunze, and Shadi Tahvildar-Zadeh for helpful discussions.
 They also thank the two referees and the editor, Herbert Spohn, for their comments. 

\newpage

\appendix
\section*{Appendix}
\addcontentsline{toc}{section}{Appendices}
\renewcommand{\thesubsection}{\Alph{subsection}}
\numberwithin{equation}{subsection}

\subsection{The radiation-reaction force term}

 In this appendix we collect the pertinent formulas needed to compute the radiation-reaction force terms.

 We write $\PiV^{\mbox{\tiny{field}}}_{\alpha,n}(t,\sV)$ as a sum of three terms, sorted by their singularities, 
\begin{alignat}{1}
\label{eq:PiDECOMPOSEDn}
\PiV^{\mbox{\tiny{field}}}_{\alpha,n}(t,\sV)
 = 
\frac{e_\alpha^2}{16\pi^2 c}
\;{\sum\limits_{k=0}^2}\; 
\frac{\boldsymbol{\pi}^{[k]}_{\boldsymbol{\xi}_n^\alpha}(t,\sV)}{\big|{\sV-\qV_n\big(t^{\mathrm{ret}}_{\boldsymbol{\xi}_n^\alpha}(t,\sV)\big)}\big|^k} ,
% \boldsymbol{\pi}_{\boldsymbol{\xi}_n}(t,\sV) 
% + \frac{\widetilde{\mathbf{Z}}_{\!\!\qV_n}(t,\sV)}{\Abs{\sV-\qV_n\big(t^{\mathrm{ret}}_{\boldsymbol{\xi}_n}(t,\sV)\big)}}  
% +  \frac{\mathbf{Z}_{\boldsymbol{\xi}_n}(t,\sV)}{\Abs{\sV-\qV_n\big(t^{\mathrm{ret}}_{\boldsymbol{\xi}_n}(t,\sV)\big)}^2} ,
\end{alignat} 
where the suffix $\boldsymbol{\xi}_n^\alpha$ indicates the vector function
 $t\mapsto\boldsymbol{\xi}_n^\alpha(t) \equiv (\qV_n^\alpha,\vV_n^\alpha,\aV_n^\alpha)(t)$.
 Here, and now dropping ${}_n$ and ${}^\alpha$ indices,
\begin{alignat}{1}
\label{eq:X0n}
\hspace{-0.5truecm}
 \boldsymbol{\pi}_{\boldsymbol{\xi}}^{[0]}(t,\sV) =
& - \varkappa^4 \frac14\left[
{\textstyle{
\frac{ \left({\nV(\qV,\sV)}_{\phantom{!\!}}-\frac1c{\vV}\right)\crprd \left({\color{black}
\frac1c{\vV}\crprd{\nV(\qV,\sV)}_{\phantom{!\!}} }\right) }{
      \bigl({1-\frac1c {\vV}\cdot\nV(\qV,\sV)}\bigr)^{\!2} }
}}\right]_{\mathrm{ret}}\\ \notag
&+ \varkappa^4\frac12\left[
{\textstyle{
\frac{ {\nV(\qV,\sV)}_{\phantom{!\!}}-\frac1c{\vV}}{ {1-\frac1c {\vV}\cdot\nV(\qV,\sV)} }
             }}\right]_{\mathrm{ret}} 
\crprd \!
\int_{-\infty}^{t^\mathrm{ret}_{\boldsymbol{\xi}}(t,\sV)}\!\!\!\!
{\vV(t')}\crprd \mathbf{K}_{\boldsymbol{\xi}}(t',t,\sV)\drm{t'} 
\\ \notag
& - \varkappa^4\frac12\left[{\textstyle{\frac{\color{black}
          \frac1c{\vV}\crprd \nV(\qV,\sV)}{ 1-\frac1c {\vV}\cdot\nV(\qV,\sV)}}}\right]_{\mathrm{ret}} 
\crprd \int_{-\infty}^{t^\mathrm{ret}_{\boldsymbol{\xi}}(t,\sV)}\!\!\!\!
 c\mathbf{K}_{\boldsymbol{\xi}}(t',t,\sV)\drm{t'} 
\\ \notag
& - \varkappa^4 \int_{-\infty}^{t^\mathrm{ret}_{\boldsymbol{\xi}}(t,\sV)} \!\!\!\!
 c\mathbf{K}_{\boldsymbol{\xi}}(t',t,\sV)\drm{t'} \crprd \int_{-\infty}^{t^\mathrm{ret}_{\boldsymbol{\xi}}(t,\sV)} \!\!\!\!
{\vV(t')}\crprd \mathbf{K}_{\boldsymbol{\xi}}(t',t,\sV)\drm{t'} 
\\ \notag
& - \varkappa^4 c\int_{-\infty}^{t^\mathrm{ret}_{\boldsymbol{\xi}}(t,\sV)} \!\!\!\! \mathrm{K}_{\boldsymbol{\xi}}(t',t,\sV)\drm{t'} 
 \int_{-\infty}^{t^\mathrm{ret}_{\boldsymbol{\xi}}(t,\sV)}  \!\!\!\!\mathrm{K}_{\boldsymbol{\xi}}(t',t,\sV) {\vV}(t')\drm{t'} 
\end{alignat}
\begin{alignat}{1}
\hspace{-0.9truecm}
 \boldsymbol{\pi}_{\boldsymbol{\xi}}^{[1]}(t,\sV) = & 
- \varkappa^2 \left[
{\textstyle{
{\color{black}
{\nV(\qV,\sV)}\frac{\left({\nV(\qV,\sV)}\crprd [{ \left({\nV(\qV,\sV)}-\frac1c{\vV}\right)\crprd \aV }]\right)\cdot\frac1c\vV}{
       c^2 \bigl({1-\frac1c {\vV}\cdot\nV(\qV,\sV)}\bigr)^{\!4} }
    }
+ {\nV(\qV,\sV)}_{\phantom{!\!}}\crprd\frac{ \left({\nV(\qV,\sV)}_{\phantom{!\!}}-\frac1c{\vV}\right)\crprd \aV }{
      2 c^2 \bigl({1-\frac1c {\vV}\cdot\nV(\qV,\sV)}\bigr)^{\!3} }
}}\right]_{\mathrm{ret}}\notag \\  \label{eq:X1n}
&- \varkappa^2\left[
{\textstyle{
{\nV(\qV,\sV)}_{\phantom{!\!}}\crprd\frac{ \left({\nV(\qV,\sV)}_{\phantom{!\!}}-\frac1c{\vV}\right)\crprd \aV }{
      c^2 \bigl({1-\frac1c {\vV}\cdot\nV(\qV,\sV)}\bigr)^{\!3} }
}}\right]_{\mathrm{ret}} \!\!
\crprd \!
\int_{-\infty}^{t^\mathrm{ret}_{\boldsymbol{\xi}}(t,\sV)}\!\!\!\!
{\vV(t')}\crprd \mathbf{K}_{\boldsymbol{\xi}}(t',t,\sV)\drm{t'} 
\\ \notag
& + \varkappa^2\left[\nV(\qV,\sV)\crprd \biggl[{\textstyle{\nV(\qV,\sV)\crprd 
\frac{\left({\nV(\qV,\sV)}_{\phantom{!\!}}-\frac1c{\vV}\right)\crprd\aV }{
      c^2\bigl({1-\frac1c {\vV}\cdot\nV(\qV,\sV)}\bigr)^{\!3} }
}}\biggr]\right]_{\mathrm{ret}} \!\!\!
\crprd\! \int_{-\infty}^{t^\mathrm{ret}_{\boldsymbol{\xi}}(t,\sV)} \!\!\!\!
 c\mathbf{K}_{\boldsymbol{\xi}}(t',t,\sV)\drm{t'} 
\\ \notag
& +  \varkappa^3 \left[\textstyle\frac{1}{{1-\frac1c {\vV}\cdot\nV(\qV,\sV)} }\right]_{\mathrm{ret}} 
 \int_{-\infty}^{t^\mathrm{ret}_{\boldsymbol{\xi}}(t,\sV)}\!\!\!\!
 \mathrm{K}_{\boldsymbol{\xi}}(t',t,\sV)\left[{\vV}({t^\mathrm{ret}_{\boldsymbol{\xi}}(t,\sV)}))+{\vV}(t')\right]\drm{t'} 
\end{alignat}
\begin{alignat}{1}
\label{eq:X2n} 
\hspace{-0.9truecm}
\boldsymbol{\pi}_{\boldsymbol{\xi}}^{[2]}(t,s) = & - \varkappa^2
\left[\textstyle\frac{1}{\bigl({1-\frac1c {\vV}\cdot\nV(\qV,\sV)}\bigr)^{\!2} }\frac1c{\vV}
{\color{black}- \Big[\!{1-\tfrac{1}{c^2}\big|\vV\big|^2}\!\Big]
\frac{ \left({\nV(\qV,\sV)}_{\phantom{!\!}}-\frac1c{\vV}\right) \crprd \left(\frac1c\vV\crprd \nV(\qV,\sV)\right) }{
      \bigl({1-\frac1c {\vV}\cdot\nV(\qV,\sV)}\bigr)^4 }}
\right]_{\mathrm{ret}} \hspace{-20pt}
\\ \notag
& +\varkappa^2 \left[\Big[\!{1-\tfrac{1}{c^2}\big|\vV\big|^2}\!\Big]\nV(\qV,\sV)\crprd {\textstyle{
\frac{ {\nV(\qV,\sV)}_{\phantom{!\!}}-\frac1c{\vV} }{
      \bigl({1-\frac1c {\vV}\cdot\nV(\qV,\sV)}\bigr)^{\!3} }
}}\right]_{\mathrm{ret}} \crprd\int_{-\infty}^{t^\mathrm{ret}_{\boldsymbol{\xi}}(t,\sV)} \!\!\!\!
 c\mathbf{K}_{\boldsymbol{\xi}}(t',t,\sV)\drm{t'} \\
\notag
&-  \varkappa^2\left[\Big[\!{1-\tfrac{1}{c^2}\big|\vV\big|^2}\!\Big]
{\textstyle{
\frac{ {\nV(\qV,\sV)}_{\phantom{!\!}}-\frac1c{\vV} }{
      \bigl({1-\frac1c {\vV}\cdot\nV(\qV,\sV)}\bigr)^{\!3} }
}}\right]_{\mathrm{ret}} 
\crprd \!
\int_{-\infty}^{t^\mathrm{ret}_{\boldsymbol{\xi}}(t,\sV)} \!\!\!\!
{\vV(t')}\crprd \mathbf{K}_{\boldsymbol{\xi}}(t',t,\sV)\drm{t'} ,
\end{alignat}
with the abbreviations $\mathrm{K}_{\boldsymbol{\xi}}, \mathbf{K}_{\boldsymbol{\xi}}$ defined as in 
\refeq{eq:KbfTILDE} and \refeq{eq:KrmTILDE}, 
%\begin{alignat}{1}
%\mathrm{K}_{\boldsymbol{\xi}}(t',t,\sV) & :=\label{eq:Krm}
%\tfrac{J_1\!\bigl(\varkappa\sqrt{c^2(t-t')^2-|\sV-\qV(t')|^2 }\bigr)}{\sqrt{c^2(t-t')^2-|\sV-\qV(t')|^2}^{\phantom{n}}},\\
%\mathbf{K}_{\boldsymbol{\xi}}(t',t,\sV) & := \label{eq:Kbf}
%\tfrac{J_2\!\bigl(\varkappa\sqrt{c^2(t-t')^2-|\sV-\qV(t')|^2 }\bigr)}{{c^2(t-t')^2-|\sV-\qV(t')|^2}^{\phantom{n}} }
% \left(\sV-\qV(t')- \vV(t')(t-t')\right),
%\end{alignat}
and where $\big|_\mathrm{ret}$ means that $\qV(\tilde{t})$, $\vV(\tilde{t})$, and $\aV(\tilde{t})$ 
are evaluated with $\tilde{t} = {t^\mathrm{ret}_{\boldsymbol{\xi}}}(t,\sV)$.

 With $\overline{\boldsymbol{\xi}}\equiv (\olqV,\vV_{\!0},\NullV)$, we thus have
\begin{alignat}{1}
\label{eq:PIminusPInullDECOMP}
&{\displaystyle\int_{B_{ct}(\qV_n(0))}\!\!
 \left(\PiV^{\mbox{\tiny{field}}}_{\alpha,n}(t,\sV)-{\PiV}^{\mbox{\tiny{field}}}_{\alpha,n}(0,\sV-\olqV_n(t))\right)\!\drm^3{s}}
=  \\ \notag
& \qquad\qquad\qquad
 \frac{e_\alpha^2}{16\pi^2 c} \;{\sum\limits_{k=0}^2}\; 
{\displaystyle\int_{B_{ct}(\qV_0)}\!\!\left(
\tfrac{\textstyle\boldsymbol{\pi}_{\boldsymbol{\xi}_n^\alpha}^{[k]}(t,\sV)}{\Abs{\sV-\qV_n\big(t^{\mathrm{ret}}_{\boldsymbol{\xi}_n}(t,\sV)\big)}^k} 
-
\tfrac{\textstyle\boldsymbol{\pi}_{\overline{\boldsymbol{\xi}}_n^\alpha}^{[k]}(t,\sV)}{\Abs{\sV-\olqV_n\big(t^{\mathrm{ret}}_{\olqV_n}(t,\sV)\big)}^k} 
\right) \drm^3{s}}.   
\end{alignat}
 Note that the acceleration $t\mapsto\aV_n(t)$ only features linearly, in the $k=1$ term.

 To carry out the integrations at r.h.s.\refeq{eq:PIminusPInullDECOMP} we in \cite{KTZonBLTP}
switch to ``retarded spherical co-ordinates'' 
with the colatitude $\vartheta$ defined with respect to $\sV - \qV (t^{\mathrm{r}})$. 
 Let $\drm\omega_r(\sV)$ denote the uniform measure on  $\partial B_r\left(\qV(t^{\mathrm{r}})\right)$.
 For any $\Csp^{1,1}$ map $t^{\mathrm{r}}\mapsto\qV(t^{\mathrm{r}})$ with $\dot\qV(t^{\mathrm{r}})=:\vV(t^{\mathrm{r}})$ 
we then have
\begin{alignat}{1}
\label{eq:COareaINTelement}\hspace{-15pt}
\drm^3{s} = 
\frac{\drm{\omega_r(\sV)}}{\abs{\nab{\Abs{\sV-\qV(t^{\mathrm{r}})}}}}
\biggl.\biggr|_{\partial B_r\left(\qV(t^{\mathrm{r}})\right)}^{}\hspace{-10pt}\drm{r}
=
 \left(1-\tfrac1c\abs{\vV(t-\tfrac1c r)}\cos\vartheta\right)
 r^2\sin\vartheta \drm{\vartheta}\drm{\varphi}\drm{r}.
\end{alignat}
 This gives us
\begin{alignat}{1}
\label{eq:XkCOarea} 
& \displaystyle\int_{B_{ct}(\qV_0)}
\biggl[
\tfrac{\textstyle\boldsymbol{\pi}_{\boldsymbol{\xi}_n}^{[k]}(t,\sV)}{\Abs{\sV-\qV_n\big(t^{\mathrm{ret}}_{\boldsymbol{\xi}_n}(t,\sV)\big)}^k}  
-
\tfrac{\textstyle\boldsymbol{\pi}_{\boldsymbol{\xi}_n^\circ}^{[k]}(t,\sV)}{\Abs{\sV-\olqV_n\big(t^{\mathrm{ret}}_{\olqV_n}(t,\sV)\big)}^k}  
\biggr]             \drm^3{s} = \\ \notag 
& \hspace{2truecm} \displaystyle  \hspace{-1truecm}\int_0^{ct}\! \int_0^{2\pi}\!\! \int_0^{\pi}\!
\Bigl[\left(1-\tfrac1c\abs{\vV_n^{}(t-\tfrac1c r)}\cos\vartheta\right)
 \boldsymbol{\pi}_{\boldsymbol{\xi}_n}^{[k]}\big(t,\sV_{\boldsymbol{\xi}_n}(r,\vartheta,\varphi)\big)\biggr.
-\\ \notag
&\hspace{3.5truecm}\biggl. \left(1-\tfrac1c\abs{\vV_{\!n}(0)}\cos\vartheta\right)
 \boldsymbol{\pi}_{\boldsymbol{\xi}_n^\circ}^{[k]}\big(t,\sV_{\olqV_n}(r,\vartheta,\varphi)\big)\Bigr]
\sin\vartheta \drm{\vartheta}\drm{\varphi}\,  r^{2-k} \drm{r} \,.
 \end{alignat}

 Next we register the following important facts about the angular integrations over the spheres 
$\partial B_r\left(\qV(t^{\mathrm{r}})\right)$ (where $\qV$ stands for either $\qV_n$ or $\olqV_n$; $\vV$ and $\aV$ similarly):
\begin{itemize}
\item 
on $\partial B_{r}(\qV(t^{\mbox{\tiny{r}}}))$ the retarded time 
$t^{\mbox{\tiny{ret}}}_{\qV}(t,\sV)  = t-\frac1c r = t^{\mbox{\tiny{r}}}$ is constant;\vspace{-5pt}
\item
the vectors $\qV(t^{\mbox{\tiny{r}}})$, $\vV(t^{\mbox{\tiny{r}}})$, and $\aV(t^{\mbox{\tiny{r}}})$ 
%, evaluated at $t^{\mbox{\tiny{r}}} = t^{\mbox{\tiny{ret}}}_{\qV}(t,\sV) = t-r/c$,
are constant during the integration over $\partial B_{r}(\qV(t^{\mbox{\tiny{r}}}))$;\vspace{-5pt}
\item
for $\sV\in\partial B_{r}(\qV(t^{\mbox{\tiny{r}}}))$, we have
$\sV = r\, \nV(\qV(t^{\mbox{\tiny{r}}}),\sV) + \qV\big(t^{\mbox{\tiny{r}}}\big)$;\vspace{-5pt}
\item
for $\sV\in\partial B_{r}(\qV(t^{\mbox{\tiny{r}}}))$ 
the unit vector $\nV(\qV(t^{\mbox{\tiny{r}}}),\sV)=\big(\sin\vartheta \cos\varphi, \sin\vartheta \sin\varphi, \cos\vartheta\big)$.
\end{itemize}
 It follows that on $\partial B_{r}(\qV(t^{\mbox{\tiny{r}}}))$ the vectors 
$\boldsymbol{\pi}_{\boldsymbol{\xi}}^{[k]}\big(t,\sV_{\qV}(r,\vartheta,\varphi)\big)$ are bounded continuous functions of $\vartheta$ and $\varphi$.
 Thus the angular integrations at r.h.s.\refeq{eq:XkCOarea} can be carried out to yield
\begin{alignat}{1}
\label{eq:XkCOareaINToverSPHERESdone} 
\hspace{-1truecm}
\displaystyle\int_{B_{ct}(\qV_0)}
\biggl[
\tfrac{\textstyle\boldsymbol{\pi}_{\boldsymbol{\xi}_n}^{[k]}(t,\sV)}{\Abs{\sV-\qV_n\big(t^{\mathrm{ret}}_{\boldsymbol{\xi}_n}(t,\sV)\big)}^k}  
-
\tfrac{\textstyle\boldsymbol{\pi}_{\boldsymbol{\xi}_n^\circ}^{[k]}(t,\sV)}{\Abs{\sV-\olqV_n\big(t^{\mathrm{ret}}_{\olqV_n}(t,\sV)\big)}^k}  
\biggr] \drm^3{s} = \!\!
\displaystyle  \int_0^{ct}\! 
\Bigl[ \widetilde{\mathbf{Z}}_{\boldsymbol{\xi}_n}^{[k]}\big(t,r\big)
-\!
 \widetilde{\mathbf{Z}}_{\boldsymbol{\xi}_n^\circ}^{[k]}\big(t,r\big)\Bigr]
 r^{2-k} \drm{r} .\hspace{-1truecm}
 \end{alignat}
 The vector functions $\widetilde{\mathbf{Z}}_{\boldsymbol{\xi}}^{[k]}(t,r)$ depend on the maps
$t'\mapsto \qV(t')$ and $t'\mapsto \vV(t')$
between $t'=0$ and $t'=t^{\mbox{\tiny{ret}}}_{\qV}(t,\sV)=t^{\mbox{\tiny{r}}}$,
through the $\drm{t'}$ integrals over the Bessel function kernels, through which they also 
depend on the initial data $\qV(0)$ and $\vV(0)$ (recall that the integrations over $t'<0$
involve an unaccelerated auxiliary motion determined by the initial data $\qV(0)$ and $\vV(0)$).
 They also depend on $t$ and $r$, explicitly through the Bessel function kernels \refeq{eq:KrmTILDE} and \refeq{eq:KbfTILDE}
(recall also the third bullet point above), and implicitly through the dependence on $t^{\mbox{\tiny{r}}} = t-r/c$ in:
(a) the vectors $\qV(t^{\mbox{\tiny{r}}})$, $\vV(t^{\mbox{\tiny{r}}})$, and (when $k=1$) $\aV(t^{\mbox{\tiny{r}}})$, and
(b) the upper limit of integration of the integrals over the Bessel function kernels. 

 Next, by a change of integration variable from $r$ to $t^{\mbox{\tiny{r}}}= t-r/c$ we eliminate the 
implicit $t$ dependence from the integrands at r.h.s.\refeq{eq:XkCOareaINToverSPHERESdone}. 
 Writing 
$\widetilde{\mathbf{Z}}_{\boldsymbol{\xi}}^{[k]}\big(t,r\big) = {\mathbf{Z}}_{\boldsymbol{\xi}}^{[k]}\big(t,t^{\mbox{\tiny{r}}}\big)$,
this yields \refeq{eq:PIminusPInullDECOMPexplicated}.

 We thus arrive at 
the ``self''-field force on the $n$-th point charge source of the MBLTP field equations,
given for $t>0$ by the negative $t$ derivative of \refeq{eq:PIminusPInullDECOMPexplicated}, viz.
\begin{alignat}{1}
\label{eq:SELFforceEXPLICIT}
& -\tfrac{16\pi^2 }{e_\alpha^2}{\textstyle\Ddt}{\displaystyle\int_{B_{ct}(\qV_n(0))}\!\!
 \left(\PiV^{\mbox{\tiny{field}}}_{\alpha,n}(t,\sV)-{\PiV}^{\mbox{\tiny{field}}}_{\alpha,n}(0,\sV-\olqV^\alpha_n(t))\right)\!\drm^3{s}}
 = \;
\\ \notag
 &- {\mathbf{Z}}_{\boldsymbol{\xi}_n}^{[2]}(t,t) + {\mathbf{Z}}_{\boldsymbol{\xi}_n^\circ}^{[2]}(t,t) \, -
\\ \notag
&   \;{\textstyle\sum\limits_{0\leq k\leq 1}}\; c^{2-k}(2-k)
\displaystyle  \int_0^{t}\! 
\Bigl[{\mathbf{Z}}_{\boldsymbol{\xi}_n}^{[k]}\big(t,t^{\mbox{\tiny{r}}}\big)
-\!
{\mathbf{Z}}_{\boldsymbol{\xi}_n^\circ}^{[k]}\big(t, t^{\mbox{\tiny{r}}}\big)\Bigr]
(t- t^{\mbox{\tiny{r}}})^{1-k} \drm{t^{\mbox{\tiny{r}}}} -
\\ \notag
&  \;{\textstyle\sum\limits_{0\leq k\leq 2}}\; c^{2-k}
\displaystyle  \int_0^{t}\! 
\Bigl[\tpddt{\mathbf{Z}}_{\boldsymbol{\xi}_n}^{[k]}\big(t,t^{\mbox{\tiny{r}}}\big)
-\!
\tpddt{\mathbf{Z}}_{\boldsymbol{\xi}_n^\circ}^{[k]}\big(t, t^{\mbox{\tiny{r}}}\big)\Bigr]
(t- t^{\mbox{\tiny{r}}})^{2-k} \drm{t^{\mbox{\tiny{r}}}} . 
\end{alignat}
\medskip
\noindent
{\sc{Remark A.1}}%\label{rem:BLTPselfFORCEa}
\emph{Thus the integrals in the second and third line at r.h.s.\refeq{eq:SELFforceEXPLICIT} are well defined Riemann integrals.}

\medskip
\noindent
{\sc{Remark A.2}} %\label{rem:BLTPselfFORCEb}
\emph{Since $t^{\mathrm{r}}$ is a mute integration variable at r.h.s.\refeq{eq:SELFforceEXPLICIT},
the $\pddt$ derivative in the third line at r.h.s\refeq{eq:SELFforceEXPLICIT} does not act on it.
 In particular, it does not act on $\qV_n(t^\mathrm{r})$, $\vV_n(t^\mathrm{r})$, or $\aV_n(t^\mathrm{r})$ in any of the
${\mathbf{Z}}_{\boldsymbol{\xi}_n}^{[k]}\big(t,t^{\mathrm{r}}\big)$.
 This leads to the important observation that the ``self'' force does not involve higher-than-second-order
time derivates of $t\mapsto\qV(t)$.}

\medskip
\noindent
{\sc{Remark A.3}} %\label{rem:BLTPselfFORCEc}
\emph{The derivative $\pddt$ also does not act on $\qV_n(t^\prime)$ or $\vV_n(t^\prime)$ in the kernels 
$\mathbf{K}_{\boldsymbol{\xi}}(t',t,\sV_\qV)$ and $\mathrm{K}_{\boldsymbol{\xi}}(t',t,\sV_\qV)$. 
 Thus the acceleration enters the ``self'' force only linearly, as $\aV_n(t^{\mathrm{r}})$, through 
${\mathbf{Z}}_{\boldsymbol{\xi}_n}^{[1]}\big(t,t^{\mathrm{r}}\big)$ in the second, 
and $\pddt{\mathbf{Z}}_{\boldsymbol{\xi}_n}^{[1]}\big(t,t^{\mathrm{r}}\big)$ in the third line at r.h.s.\refeq{eq:SELFforceEXPLICIT}.
 This gives rise to the Volterra integral equation for the accelerations, mentioned in 2.3.}

 There is no purely light-like contribution to the third line at r.h.s.\refeq{eq:SELFforceEXPLICIT}.
 The $\pddt$ in the third line at r.h.s.\refeq{eq:SELFforceEXPLICIT}
acts only on the explicit $t$-dependence in $\mathbf{K}_{\boldsymbol{\xi}}$ and in $\mathrm{K}_{\boldsymbol{\xi}}$ 
in the mixed light- \&\ time-like and purely time-like contributions to 
${\mathbf{Z}}_{\boldsymbol{\xi}_n}^{[k]}\big(t,t^{\mathrm{r}}\big)$.
The identity 
\begin{equation} \label{eq:BESSELderivativeID}
\frac1x \frac{\drm}{\drm x}\frac{J_\nu(x)}{x^\nu} 
= - \frac{J_{\nu+1}(x)}{x^{\nu+1}} 
\end{equation} 
(see \S 10.6(ii) in \cite{DLMF}) yields these derivatives easily as
\begin{alignat}{1}
\hspace{-1truecm}
{\textstyle\pddt}\mathrm{K}_{\boldsymbol{\xi}}(t',t,\sV_\qV)
 = & \label{eq:KrmPRIME}
-\varkappa c^2 \tfrac{J_2\!\bigl(\varkappa\sqrt{c^2(t-t')^2-|\sV_\qV-\qV(t')|^2 }\bigr)}
{{c^2(t-t')^2-|\sV_\qV-\qV(t')|^2}^{\phantom{n}}} (t-t'),\\
\hspace{-1truecm}
{\textstyle\pddt}\mathbf{K}_{\boldsymbol{\xi}}(t',t,\sV_\qV) 
 =& 
-  \varkappa c^2 
\tfrac{J_3\!\bigl(\varkappa\sqrt{c^2(t-t')^2-|\sV_\qV-\qV(t')|^2 }\bigr)}{(c^2(t-t')^2-|\sV_\qV-\qV(t')|^2)^{3/2} }
(t-t') \left(\sV_\qV-\qV(t')- \vV(t')(t-t')\right) \!\!\!
\label{eq:KbfPRIME} \\
&- \vV(t') \tfrac{J_2\!\bigl(\varkappa\sqrt{c^2(t-t')^2-|\sV_\qV-\qV(t')|^2 }\bigr)}{{c^2(t-t')^2-|\sV_\qV-\qV(t')|^2}^{\phantom{n}} },
\notag
\end{alignat}
where $\sV_\qV = r\nV +\qV(t^\mathrm{r})$.

%===========================================================
\newpage
{\color{black}
\subsection{A factorization lemma}

 For the following lemma and its proof we may ignore the time variable, 
and we abbreviate $(\pV^{},\qV^{})$ by $x\in\Rset^6$ and $(\tilde\pV,\tilde\qV)$ by $y\in\Rset^6$.
 We write 
\begin{equation}
\uli{\triangle}^{(1)}_{N}(x)
 = \label{eq:ONEptEMPmeas}  
\ 
 \textstyle{\frac{1}{N} \sum\limits_{1\leq j\leq N }} \delta_{x_j}(x)
\end{equation}
\begin{equation}
\uli{\triangle}^{(2)}_{N}(x,y)
 = \label{eq:TWOptEMPmeas}  
\ 
\textstyle{\frac{1}{N(N-1)}\sum\!\!\!\!\sum\limits_{\hspace{-.4truecm}\genfrac{}{}{0pt}{2}{1\leq j\neq k \leq N }{}}} \delta_{x_j}(x)\delta_{x_k}(y)
\end{equation}
 For the sake of concreteness, we work with the following Kantorovich-Rubinstein distance between normalized measures on $\Rset^d$, 
where $d=6$ or $12$; viz.
\begin{equation}
\dist{\mu,\nu} := \sup \left\{ \Big|\int \ell (\xi) \drm(\mu-\nu)\Big| : \ell \in \Csp^{0,1}_b, \Lip(\ell)\leq 1, |\ell|\leq 1\right\}
\end{equation}

\noindent
{\sc{Lemma B.1}} %\label{lem:empTOprod}
\emph{If $\dist{\uli{\triangle}^{(1)}_{N}(x), f(x)}
\stackrel{N\to\infty}{\longrightarrow} 0$,
       then
$\dist{\uli{\triangle}^{(2)}_{N}(x,y), {f}(x) {f}(y)}\stackrel{N\to\infty}{\longrightarrow}~0$.}

\noindent\emph{Proof}:
 Let $\ell(x,y)$ be a bounded Lipschitz function. 
 Then $\ell(x,x)$ and $\int \ell(x,y)f(y)\drm{y}$ are also bounded Lipschitz functions.
 By elementary algebra,
\begin{alignat}{1}\hspace{-1truecm}
\iint \ell(x,y)\uli{\triangle}^{(2)}_{N}(x,y)\drm{x}\drm{y} = & 
\textstyle{\frac{1}{N(N-1)}\sum\!\!\!\!\sum\limits_{\hspace{-.35truecm}\genfrac{}{}{0pt}{2}{1\leq j\neq k \leq N }{}}}\ell(x_j,x_k) \\
\label{twoPTid}
= & \left(1 + \tfrac{1}{N-1}\right)
\textstyle{\frac{1}{N^2}\sum\!\!\!\!\sum\limits_{\hspace{-.4truecm}\genfrac{}{}{0pt}{2}{1\leq j, k \leq N }{}}}\!\!\ell(x_j,x_k) 
- \textstyle\frac{1}{N(N-1)}\sum\limits_{1\leq j\leq N}\!\!\ell(x_j,x_j) \hspace{-10pt}
\end{alignat}
 For the first term at rhs(\ref{twoPTid}) we have
\begin{alignat}{1}\hspace{-1truecm}
{\textstyle{\frac{1}{N^2}\sum\!\!\!\!\sum\limits_{\hspace{-.35truecm}\genfrac{}{}{0pt}{2}{1\leq j, k \leq N }{}}}}\!\!\ell(x_j,x_k) 
= 
\iint \ell(x,y)\uli{\triangle}^{(1)}_{N}(x) \uli{\triangle}^{(1)}_{N}(y)\drm{x}\drm{y} 
= 
\iint \ell(x,y)f(x) f(y)\drm{x}\drm{y} + o(1),
\label{twoPTidA}
\end{alignat}
while for the second term  at rhs(\ref{twoPTid}) we have
\begin{alignat}{1}\hspace{-1truecm} \label{twoPTidB}
{ \textstyle\frac{1}{N}\sum\limits_{1\leq j\leq N}}\!\!\ell(x_j,x_j)
= 
\int \ell(x,x) \uli{\triangle}^{(1)}_{N}(x)\drm{x}
=
\int \ell(x,x) f(x) \drm{x} + o(1);
\end{alignat}
here, $o(1)\to 0$ as $N\to\infty$.
 Thus
\begin{alignat}{1}\hspace{-1truecm}
\iint \ell(x,y)\uli{\triangle}^{(2)}_{N}(x,y)\drm{x}\drm{y} = 
\iint \ell(x,y) f(x) f(y)\drm{x}\drm{y} 
+ o(1).
\end{alignat}
\QED

Under regularity assumptions on $f$, the $o(1)$ can be improved to $\cO(1/N)$.
}
\vfill
\vfill
\newpage

                %%%%%%%%   BIBLIOGRAPHY  %%%%%%%%%%%

\end{document}